\documentclass[10pt]{article}

\usepackage[margin=2cm]{geometry}
\usepackage{amsfonts}
\usepackage{graphicx} 
\usepackage{subfig}
\usepackage{cite}
\usepackage{caption}
\usepackage{amsmath}
\usepackage{amssymb}
\usepackage{color}
\usepackage{soul}
\usepackage[normalem]{ulem}
\usepackage{authblk}  
\usepackage[utf8]{inputenc}
\usepackage{epsfig}
\usepackage{float} 
\usepackage{tikz}
\usetikzlibrary{patterns}
\usepackage[toc,page]{appendix}

\definecolor{DarkGreen}{RGB}{0,215,0}

\title{\textbf{Equatorial timelike circular orbits\\  around  generic ultracompact objects}}
\author[1]{Jorge F. M. Delgado\footnote{jorgedelgado@ua.pt}}
\author[1]{Carlos A. R. Herdeiro\footnote{herdeiro@ua.pt}} \author[1]{Eugen Radu\footnote{eugen.radu@ua.pt}}
\affil[1]{\normalsize Departamento de Matemática da Universidade de Aveiro and 

Center for Research and Development in Mathematics and Applications -- CIDMA

Campus de Santiago, 3810-183 Aveiro, Portugal}

\begin{document}

%
%
%
%
%
%
%
%
%
%
%

\date{\today}
\maketitle

\begin{abstract}
\normalsize

For a stationary, axisymmetric, asymptotically flat, ultra-compact  [\textit{i.e.} containing light-rings (LRs)] object, with a $\mathbb{Z}_2$ north-south symmetry fixing an equatorial plane, we establish that the structure of timelike circular orbits (TCOs) in the vicinity of the equatorial LRs, for either rotation direction, depends exclusively on the \textit{radial} stability of the LRs. Thus, an unstable LR delimits a region of unstable TCOs (no TCOs) radially above (below) it; a  stable LR delimits a region of  stable TCOs (no TCOs) radially below (above) it.
Corollaries  are discussed for both horizonless ultra-compact objects and black holes. We illustrate these results with a variety of exotic stars  examples and non-Kerr black holes, for which we also compute the efficiency associated with converting gravitational energy into radiation by a material particle falling under an adiabatic sequence of TCOs. For most objects studied, it is possible to obtain efficiencies larger than the maximal efficiency of Kerr black holes, $i.e.$ larger than $42\%$.

\end{abstract}

\tableofcontents

\section{Introduction}
The discovery that \textit{quasars} are powerful extragalactic radio sources~\cite{Schmidt:1963wkp} raised the intriguing question of how their luminosities are produced. Eventually, supermassive black holes (BHs) emerged as the widely acknowledged engines for such extreme energy outputs~\cite{Begelman:1984mw}, due to their deep gravitational potential wells. This was a turning point in the history of BHs, which slowly started to be considered as realistic physical objects by the wider astrophysics community.

The paradigmatic General Relativity  BH, described by the Kerr metric~\cite{Kerr:1963ud}, has an equatorial innermost stable circular orbit (ISCO), below which material particles trapped in the BH's potential well are expected to plunge into the horizon.\footnote{This in contrast with the analogue Keplerian problem, wherein stable circular orbits are admissible at any radius.} Thus, computing the energy per unit mass of a particle at the ISCO, $E_{\rm ISCO}$, gives an  estimate of the rest mass to radiation energy conversion by the BH.  The rationale is that a particle in, say, an equatorial thin accretion disk, moves towards smaller and smaller stable timelike circular orbits (TCOs) losing angular momentum (due to turbulence in the disk)  and converting its energy into radiation (by heating up), starting off at a large radius until it reaches the ISCO, from which  it plunges into the BH. Thus, the \textit{efficiency}
\begin{equation}
\epsilon\equiv 1 - E_{\rm ISCO} \ ,
\label{Eq:Efficiency}
\end{equation}
provides an estimate of the energy conversion into radiation by particles spiralling down the BH's potential well.\footnote{This simple estimate ignores the energy conversion during the plunge.} The first mention of efficiency was done by Shakura and
Syunyaev \cite{Syunyaev:1986zz}.
For a Kerr BH, the efficiency increases monotonically as one increases its spin. For the non-spinning case (Schwarzschild), $\epsilon\sim 5.7\%$, whereas for the extremal Kerr case  it reaches $\epsilon\sim 42\%$~\cite{hobson2006general,thorne2000gravitation}. Such dramatic rest mass to radiation energy conversion, well above that  observed in typical nuclear reactions (which is smaller than 1\%), explains why BHs could source powerful luminosities like those observed in quasars.

The estimates just quoted for  $\epsilon$  rely  on the structure of stable TCOs around Kerr BHs. How does this structure change for more generic BHs or even for horizonless compact objects?  This is a timely question, in view of the ongoing BH  astrophysics precision era, triggered by gravitational wave detections~\cite{Abbott:2016blz,LIGOScientific:2018mvr}, horizon scale electromagnetic observations~\cite{Akiyama:2019fyp,Akiyama:2019cqa,Akiyama:2019eap,Akiyama:2021qum} amongst other observational developments that impact on our knowledge of strong gravity systems. Within the goal of testing the \textit{Kerr hypothesis}, $i.e.$ that (near equilibrium) astrophysical BHs are well described  by the Kerr metric, it becomes instructive to consider more general models of BHs, motivated by beyond General Relativity gravitational theories or beyond the standard model of particle physics matter models, as well as consider their phenomenology,  in particular concerning $\epsilon$.

In this paper, we shall investigate the structure of TCOs around a generic class of equilibrium BHs or even horizonless, but sufficiently compact ($i.e.$ \textit{ultra}-compact), objects that could imitate BHs in some observables. For this purpose, we start off from two recent theorems on the existence and structure of \textit{light rings} (LRs), $i.e.$, null circular orbits around compact objects. Firstly, it was shown by using a topological argument that for a stationary, axisymmetric, asymptotically flat, 4-dimensional horizonless compact object that can be smoothly deformed into flat spacetime, LRs always come in pairs, one being stable and the other unstable \cite{Cunha:2017qtt}. Secondly, an adaptation of the same sort of topological argument, established that stationary, axisymmetric, asymptotically flat, 4-dimensional BHs always have, at least, one (unstable) LR \cite{Cunha:2020azh}.  With this starting point, in the first part of this paper, we shall show that for any given ultra-compact object, \textit{i.e.} any object with LRs, possessing also a (north-south) $\mathbb{Z}_2$ symmetry, the structure of the LRs determines, to a large extent (but not fully), the structure of the TCOs on its equatorial plane (Section 2 and 3). This allows us to establish a simple picture, identifying a small set of building blocks, whose combinations compose the structure of the equatorial TCOs around a generic equilibrium ultra-compact object - $cf.$ Figs.~\ref{Fig:IllusISCO}-\ref{Fig:CompIllustBHs} below. We note that we shall
 study the structure of TCO by looking into the \textit{radial} stability of the orbits. 
A more complete study of the full stability shall be left for a subsequent work.

The study of the full stability of TCOs has been already considered in the literature for the case of static (spherical or axisymmetric) spacetimes. In particular, Vieira \textit{et. al.} \cite{Vieira:2017zau} (see also the references therein) have shown  that, for such Ricci-flat spacetimes,  the sum of the radial and angular epicyclic frequencies squared on the equatorial plane,  measured by an observer at infinity, vanishes at the LRs. This implies that, for either an unstable or a stable LR, in its adjacent region supporting TCOs one of the epicyclic frequencies squared must be negative, leading always to an unstable region of TCOs.
Dropping the Ricci flatness assumption, Vieira \textit{et. al.} \cite{Vieira:2017zau} were still able to show that the aforementioned sum of the radial and angular epicyclic frequencies squared is always positive if the Strong Energy Condition is obeyed.

Then, in Section 4, we illustrate these generic structures of TCOs by considering a sample of models of alternative BHs (to the Kerr solution) and horizonless ultra-compact objects, also computing  their efficiency.  These generalised models unveil an ambiguity related to the proper definition of the efficiency parameter for objects with a  more complex structure of TCOs, that we shall discuss. Amongst the explicit examples considered, there are several family of bosonic scalar and vector stars~\cite{SCHUNCK1998389,PhysRevD.56.762,Delgado:2016jxq,Guerra:2019srj,Delgado:2020udb,Brito:2015pxa,Herdeiro:2017fhv,Minamitsuji:2018kof}, as well as two different families of ``hairy" BHs~\cite{Delgado:2020hwr,Sotiriou:2014pfa,Delgado:2020rev}. In Section 5 we present a closing discussion about our results. Throughout this paper we shall use geometrized units, $G = c = \hbar = 1$.

\section{Circular geodesics on the equatorial plane}
\label{Sec:Sec2}

We assume a stationary, axi-symmetric, asymptotically flat, 1+3 dimensional spacetime, $(\mathcal{M},g)$, describing an ultra-compact object.   $(\mathcal{M},g)$ may, or may not, have an event horizon.  Our description is theory agnostic: we do not assume  $(\mathcal{M},g)$ solves any particular model. 

Let the two Killing vectors  associated to stationarity and axi-symmetry be, respectively, $\{\eta_{1},\eta_2\}$. Then, a theorem by Carter guarantees that (due to asymptotic flatness), $[\eta_1,\eta_2]=0$, $i.e.$ the Killing vector fields commute~\cite{Carter:1970ea}. Consequently, a coordinate system adapted to \textit{both} Killing vectors can be chosen: $(t,r,\theta,\varphi)$, such that $\eta_1=\partial_t,\eta_2=\partial_\varphi$. In addition, we assume that the metric: $(i)$ admits a north-south $\mathbb{Z}_2$ symmetry; and $(ii)$ is circular. For asymptotically flat spacetimes, circularity implies that the geometry possesses a 2-space orthogonal to $\{\eta_{1},\eta_2\}$ ($c.f.$ theorem 7.1.1 in~\cite{Wald:1984rg}).  Thus $g$ admits the discrete symmetry  $(t,\varphi)\to (-t,-\varphi)$.

The spherical-like coordinates ($r,\theta$) in the orthogonal 2-space are assumed  to be orthogonal (which amounts to a gauge choice). If an  event horizon exists, another  gauge choice guarantees the horizon is located at a constant (positive) radial coordinate $r=r_H$; thus the exterior region is  $r_H<r<\infty$. If not, the radial coordinate spans $\mathbb{R}^+_0$ ($0\leqslant r<\infty)$. Under such choices, $g_{r\theta}=0$, $g_{rr}>0$ and $g_{\theta\theta}>0$ (outside the possible horizon). The $(r,\theta)$ coordinates match the standard spherical coordinates asymptotically ($r\to \infty$); thus, $\theta\in [0,\pi]$, $\varphi\in[0,2\pi[$ and $t\in]-\infty,+\infty[$. The rotation axis, $i.e.$ the set of fixed points of $\eta_2$, is  $\theta=\{0,\pi\}$; the equatorial plane, $i.e.$ the set of fixed points of the $\mathbb{Z}_2$ symmetry, is  $\theta=\pi/2$. Outside the possible horizon, causality requires $g_{\varphi\varphi}\geqslant 0$.  Thus, our  generic metric, which has signature $(-,+,+,+)$, is\footnote{In the following  we shall consider that the radial coordinate is a faithful measure of the distance to the central object. That is, that as r increases then, say, the circunferential radius of equatorial orbits of $\eta_2$ also increases.} 
\begin{equation}
ds^2=g_{tt}dt^2+2g_{t\varphi}dtd\varphi +g_{\varphi\varphi}d\varphi^2+g_{rr}dr^2+g_{\theta\theta}d\theta^2 \ .
\label{metric1}
\end{equation}
Observe that outside a possible horizon, wherein the coordinate system is valid,
\begin{equation}
B(r,\theta)\equiv g_{t\varphi}^2 - g_{tt} g_{\varphi\varphi} > 0 \ ,
\label{Bdef}
\end{equation}
which follows from the condition ${\rm det}(-g)>0$ together with a positive signature for the $(r,\theta)$-sector of the metric.
Another combination of interest, as it will become clear below, is,
\begin{equation}
	C(r,\theta) \equiv (g_{t\varphi}')^2 - g_{tt}' g_{\varphi\varphi}' \ ,
\label{Cdef}
\end{equation}
where the prime denotes the derivative \textit{w.r.t} the radial coordinate. As we shall see, there are ultra-compact objects for which this quantity becomes negative in the domain of outer communication. This impacts in the structure of circular geodesics.

Test  particle motion in the generic geometry \eqref{metric1} is ruled by the effective Lagrangian (dots denote derivatives with respect to an affine parameter, which is proper time in the timelike case), 
\begin{equation}
	2\mathcal{L} = g_{\mu\nu} \dot{x}^\mu \dot{x}^\nu = \xi~, 
\end{equation}
where  $\xi = -1, 0, +1$ for timelike, null and spacelike geodesics, respectively.
The equatorial plane is a totally geodesic submanifold, wherein the effective Lagrangian simplifies to:
\begin{equation}
\label{gen}
	2\mathcal{L} = g_{tt}(r,\theta=\pi/2) \dot{t}^2 + 2 g_{t\varphi}(r,\theta=\pi/2) \dot{t} \dot{\varphi} + g_{rr}(r,\theta=\pi/2) \dot{r}^2 + g_{\varphi \varphi}(r,\theta=\pi/2) \dot{\varphi}^2 = \xi \ .
\end{equation}
Dropping (for notation ease) the explicit radial dependence of the metric functions,  and introducing the two integrals of motion associated to the Killing vectors, the energy, $E$, and the angular momentum, $L$,
\begin{equation}
	-E \equiv  g_{t\mu}\dot{x}^\mu = g_{tt} \dot{t} + g_{t\varphi} \dot{\varphi}~, \hspace{10pt} L \equiv g_{\varphi \mu} \dot{x}^\mu = g_{t\varphi} \dot{t} + g_{\varphi\varphi} \dot{\varphi} \ ,
\end{equation}
 the Lagrangian can be recast as
\begin{equation}
	2\mathcal{L} = - \frac{A(r,E,L)}{B(r)} + g_{rr} \dot{r}^2 = \xi \ ,
\end{equation}
where 
\begin{equation}
A(r,E,L)\equiv g_{\varphi\varphi} E^2 + 2 g_{t\varphi} E L + g_{tt} L^2 \ ,
\end{equation}
and $B(r)$ is the function in eq.~\eqref{Bdef} restricted to $\theta=\pi/2$.
This suggests introducing an effective potential $V_\xi(r)$ as,
\begin{equation}
	V_\xi(r) \equiv g_{rr} \dot{r}^2 = \xi + \frac{A(r,E,L)}{B(r)} \ .
\end{equation}
Then, a particle follows a circular orbit at $r=r^{\rm cir}$ iff the following two conditions are simultaneously obeyed throughout the orbit:
\begin{equation}
V_\xi(r^{\rm cir}) = 0 \ \  \ \ \Leftrightarrow \ \  \ \  A(r^{\rm cir},E,L)= -\xi B(r^{\rm  cir})  \ ,
\label{circon1}
\end{equation} 
and
\begin{equation}
V_\xi'(r^{\rm cir}) = 0 \ \  \ \ \Leftrightarrow \ \  \ \   A'(r^{\rm cir},E,L)= -\xi B'(r^{\rm  cir})  \ ,
\label{circon2}
\end{equation} 
where prime denotes radial derivative and we have used~\eqref{circon1} to obtain  the last equation in~\eqref{circon2}. Moreover, the \textit{radial}\footnote{Henceforth, all mentions to stability shall be understood as radial stability.} stability of such a circular orbit is determined by the sign of $V_{\xi}''(r^{\rm cir})$,  which  reads, upon using~\eqref{circon1} and~\eqref{circon2}:
\begin{equation}
V_{\xi}''(r^{\rm cir})=\frac{A''(r^{\rm cir},E,L)+\xi B''(r^{\rm cir})}{B(r^{\rm cir})} \ .
\label{stability}
\end{equation}
Then,
\begin{equation}
V_{\xi}''(r^{\rm cir})>0 \ \Leftrightarrow \ {\rm \ unstable} \ ; \qquad V_{\xi}''(r^{\rm cir})<0 \ \Leftrightarrow \ {\rm \ stable} \ .
\end{equation}
As such the transition between  stable and unstable circular orbits will be determined by $V_{\xi}''=0$. In the Kerr family, for TCOs and for each rotation direction, there is  only one radius for which $V_{-1}''=0$. For a generic ultra-compact object there may be  more solutions. Thus we  define the location of the:
\begin{itemize}
	\item Marginally Stable Circular Orbit (MSCO)
\begin{equation}
V_{-1}''(r^{\rm MSCO})=0 \  \hspace{10pt} \wedge \hspace{10pt} V_{-1}'''(r^{\rm MSCO}) < 0  \ .
\label{isco1}
\end{equation}
\end{itemize}
The MSCO should be understood as the stable circular orbit with the smallest radius that is continuously  connected to spatial infinity by a set of stable TCOs.
For objects that only have one solution satisfying the condition \eqref{isco1} (such as the Kerr case), MSCO corresponds to the well-known innermost stable circular orbit (ISCO). However, for more generic (non-Kerr) objects, a more intricate structure of TCOs may be present, with other regions of stable TCOs that are \textit{not} continuously connected to spatial infinity by a set of stable TCOs. In such cases, we can define the ISCO, which will be different from the MSCO.

To motivate the definition of the latter, we observe that along circular geodesics, the angular velocity  (as measured by an observer at infinity) is
\begin{equation}
\Omega=\frac{d\varphi}{dt}=\frac{\dot{\varphi}}{\dot{t}}=-\frac{Eg_{t\varphi}+Lg_{tt}}{Eg_{\varphi\varphi}+Lg_{t\varphi}} \ .
\label{angvel}
\end{equation}
If $\Omega$ is real, circular orbit are possible (timelike, null or spacelike). Then,  in a stationary, but not static, spacetime one distinguishes between prograde/retrograde orbits, which are co-rotating/counter-rotating with the spacetime. The angular velocity, energy and angular momentum of the former [latter] are denoted as $(\Omega_+,E_+,L_+)$ $[(\Omega_-,E_-,L_-)]$ and depend on $r^{\rm cir}$. If, however,
\begin{equation}
C(r)<0 \ , 
\label{noco}
\end{equation}
where, $C(r)$ is the function in eq.~\eqref{Cdef} restricted to $\theta=\pi/2$, then $\Omega$ is not real and no equatorial circular geodesics exist (of any causal character) -- \textit{cf.} eq. \ref{Eq:AngVelTimeLike} and eq. \ref{Eq:AngVelSpaceLike} on Appendix \ref{App:AppendixA} below. This possibility, if it occurs, typically arises close to the centre of the ultra-compact object as shown in the examples below. Then, an ISCO could emerge which is different from the MSCO defined above.
We thus define the location of the:
\begin{itemize}
	\item ISCO as the smallest $r$ for which
\begin{equation}
C(r^{\rm ISCO})=0 \ , 
\label{isco21}
\end{equation}
in case this corresponds to a TCO \textit{and} occurs in the domain of outer communication; or else, the smallest $r$ for  which
\begin{equation}
V_{-1}''(r^{\rm ISCO})=0 \  \hspace{10pt} \wedge \hspace{10pt} V_{-1}'''(r^{\rm ISCO}) < 0  \ ,
\label{isco22}
\end{equation}
in case there is more than one radial solution of~\eqref{isco22}. The ISCO is  determined by either~\eqref{isco21} or~\eqref{isco22}, whatever is smaller.
\end{itemize}
Below we shall give examples wherein an ISCO arises from~\eqref{isco21} and other examples where it arises  from~\eqref{isco22} (and ISCO $\neq$ MSCO).

\subsection{TCOs}

For timelike particles, $\xi = -1$,  condition~\eqref{circon1} together with~\eqref{angvel},  determine the energy and angular  momentum for circular orbits in terms  of the angular velocity as
\begin{equation}\label{Eq:EnergyAngMomTimelike}
	E_{\pm} = - \frac{g_{tt} + g_{t\varphi} \Omega_\pm}{\sqrt{\beta_\pm}}\Big|_{r^{\rm cir}}~, \qquad  L_\pm = \frac{g_{t\varphi} + g_{\varphi\varphi} \Omega_\pm}{\sqrt{\beta_\pm}}\Big|_{r^{\rm cir}} \ ,  
\end{equation}
where we have defined 
\begin{equation}
\beta_\pm \equiv (-g_{tt} - 2 g_{t\varphi} \Omega_\pm - g_{\varphi\varphi} \Omega_\pm^2)\Big|_{r^{\rm cir}}=- A(r^{\rm cir}, \Omega_\pm, \Omega_\pm) \ .
\label{betapm}
\end{equation}
Then, the remaining condition~\eqref{circon2} yields $\Omega_\pm$ in terms of the derivatives of the  metric functions at $r^{\rm cir}$
\begin{equation}\label{Eq:AngVelTimeLike}
	\Omega_\pm = \left[ \frac{ -g_{t\varphi}' \pm \sqrt{C(r)} }{g_{\varphi\varphi}'} \right]_{r^{\rm cir}} \ .
\end{equation}
This  confirms that the angular velocity ceases to be real when $C(r)<0$.

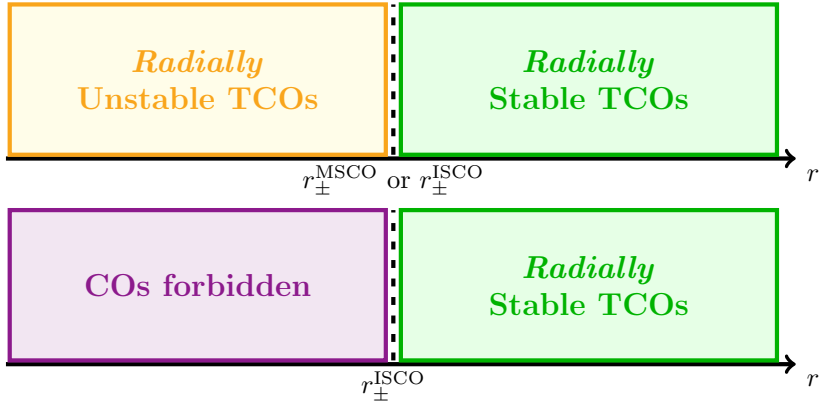
\begin{figure}[h!]
	\centering
	\begin{tikzpicture}
		\draw[ultra thick,->] (-0.05,-0.05) -- (10.45,-0.05) node[anchor=north west]{$r$};
		\filldraw[yellow!10, draw=yellow!20!orange,ultra thick] (0,0) rectangle (5,2) node[pos=0.5,yellow!20!orange,align=center,font=\large\bfseries]{\textit{Radially}\\ Unstable TCOs};
		\draw[dash pattern=on 3pt off 4pt,ultra thick] (5.1,0) -- (5.1,2);
		\filldraw[green!10, draw=green!70!black,ultra thick] (5.2,0) rectangle (10.2,2) node[pos=0.5,green!70!black,align=center,font=\large\bfseries]{\textit{Radially}\\ Stable TCOs};
		\node at (5.1,-0.35){$r_\pm^{\rm MSCO}$ or $r_\pm^{\rm ISCO}$};
	\end{tikzpicture}
	\begin{tikzpicture}
		\draw[ultra thick,->] (-0.05,-0.05) -- (10.45,-0.05) node[anchor=north west]{$r$};
		\filldraw[violet!10, draw=violet!90!white,ultra thick] (0,0) rectangle (5,2) node[pos=0.5,violet!90!white,font=\large\bfseries]{COs forbidden};
		\draw[dash pattern=on 3pt off 4pt,ultra thick] (5.1,0) -- (5.1,2);
		\filldraw[green!10, draw=green!70!black,ultra thick] (5.2,0) rectangle (10.2,2) node[pos=0.5,green!70!black, align=center,font=\large\bfseries]{\textit{Radially}\\ Stable TCOs};
		\node at (5.1,-0.35){$r_\pm^{\rm ISCO}$};
	\end{tikzpicture}
	\caption{Structure of the equatorial TCOs around the MSCO and ISCO. (Top panel) The MSCO is determined by the largest radius at the threshold of the  stability condition, $cf.$~\eqref{isco1}.  In principle this equation can have several radial solutions, so that the ISCO could also be determined by the smallest radius of the same condition, $cf.$~\eqref{isco22}. As we shall see in an example below, the latter could be at the origin, so that the yellow region of  unstable TCOs is absent. (Bottom panel) The ISCO can, alternatively, be determined by threshold condition for the absence of any circular orbits, $cf.$~\eqref{isco21}.}
	\label{Fig:IllusISCO}
\end{figure}

Asymptotically, TCOs are essentially Keplerian, and thus stable. Then, as already anticipated in the previous subsection, two important orbits amongst the TCOs emerge: the MSCO and the ISCO. For Kerr-like objects, both orbits are one and the same, and they lies at the threshold of the  stability condition~\eqref{isco1},  $V_{-1}''(r^{\rm MSCO}) = 0$, \textit{and} that it is continuously connected by stable TCOs to spatial infinity. Thus, it is determined by
%
\begin{equation}
	( g_{\varphi\varphi}'' E_\pm^2 + 2 g_{t\varphi}'' E_\pm L_\pm + g_{tt}'' L_\pm^2)\Big|_{r^{\rm MSCO}} = (g_{t\varphi}^2 - g_{tt}g_{\varphi\varphi})''\Big|_{r^{\rm MSCO}}  \ .
\end{equation}
In generic ultra-compact objects, however, and as illustrated in the examples below, there may be further disconnected regions with  stable TCOs closer to the centre of the compact object. In  particular, rather than ending at a transition to a region of  unstable TCOs, they can end at a limiting orbit below which \textit{no} circular geodesics are possible (with $C(r)<0$). 
Thus, the ISCO can occur at the threshold of this region, which is given by eq.~\eqref{isco21}:
%
\begin{equation}
	C(r^{\rm ISCO}) = [(g_{t\varphi}')^2 - g_{tt}' g_{\varphi\varphi}']_{r^{\rm ISCO}} = 0~, \hspace{10pt} \text{and} \hspace{10pt} V_{-1}''(r^{\rm ISCO} + |\delta r|) < 0 \ ,
\end{equation}
where $|\delta r| \ll 1$. An illustration of the structure of TCOs around the MSCO and ISCO is shown in Fig. \ref{Fig:IllusISCO}. We remark that below we will see ultra-compact objects where both MSCO and ISCO are present.

\subsection{LRs}

For lightlike particles, $\xi = 0$, circular orbits are LRs. Condition~\eqref{circon1} is a quadratic equation for the \textit{inverse impact parameter}, 
\begin{equation}
\sigma_\pm \equiv \frac{ E_\pm}{L_\pm},
\end{equation}
$i.e.$ it reads
\begin{equation}
	A(r^{\rm LR},\sigma_\pm,\sigma_\pm)=\left[g_{\varphi\varphi}\sigma_\pm^2 + 2 g_{t\varphi} \sigma_\pm + g_{tt} \right]_{\rm LR}= 0 \ ,
\label{LRcon1}
\end{equation}
 with the solutions (for prograde, $\sigma_+$, and retrograde, $\sigma_-$, LRs),
\begin{equation}
	\sigma_\pm =\left[  \frac{ -g_{t\varphi} \pm \sqrt{g_{t\varphi}^2 - g_{tt} g_{\varphi\varphi} }} {g_{\varphi\varphi}}\right]_{\rm LR} \ .
\label{sigmapm}
\end{equation}
The second condition~\eqref{circon2}, on the other hand, yields
\begin{equation}\label{Eq:LRsEquation}
	\left[g_{\varphi\varphi}' \sigma_\pm^2 + 2 g_{t\varphi}' \sigma_\pm + g_{tt}' \right]_{\rm LR}= 0 \ .
\end{equation}
This  determines LR's radial coordinate  and cannot  be solved if~\eqref{noco} holds.
The  stability of the LRs is evaluated by checking the sign of $V_0''(r^{\rm LR})$ given by~\eqref{stability}; explicitly
\begin{equation}\label{Eq:StabilityLRs}
	V_0''(r^{\rm LR}) = L^2_\pm \left[ \frac{g_{\varphi\varphi}'' \sigma_\pm^2 + 2 g_{t\varphi}'' \sigma_\pm + g_{tt}''}{g_{t\varphi}^2 - g_{tt} g_{\varphi\varphi}}\right]_{\rm LR} \ .
\end{equation}
The sign is determined by the numerator; if it is positive (negative) the motion is  unstable (stable).

\section{TCOs in the vicinity of LRs}

We now assume the existence of a LR (which, from the last section, requires $C(r_{\rm LR})\geqslant 0$).\footnote{The case $C(r_{\rm LR})= 0$ is rather special; although it may be realized in the examples  below it corresponds to a zero measure set in the  space of solutions. We will further comment on it below, but for now, we shall  assume the generic case $C(r_{\rm LR})> 0$.} Then, we wish to determine if TCOs exist in its immediate neighbourhood and whether they are  stable or unstable. 

\subsection{Allowed region}

First, we connect the description of timelike and null orbits. The connection amounts to observe that LRs are determined by 
\begin{equation}
	\beta_\pm\big|_{\rm LR}  = 0 \ , \qquad {\rm and \ noting \  that}  \quad  \Omega_\pm\big|_{\rm LR}=  \sigma_\pm  \ .
\end{equation}
Indeed, from~\eqref{betapm}, the condition $\beta_\pm = 0$ becomes equivalent to~\eqref{LRcon1} and~\eqref{Eq:LRsEquation} is solved by virtue of~\eqref{Eq:AngVelTimeLike}.

The function $\beta_\pm$ will guide us in the connection between LRs and TCOs.\footnote{This function can be regarded as proportional to the mass squared of the particle along the corresponding circular orbit; thus  it is positive, zero and negative, for TCOs, null circular orbits and spacelike circular orbits.}  From the continuity of $\beta_\pm$ -- see Appendix \ref{App:AppendixA} for more details --  one expects that (generically) in the neighbourhood of the LR ($r$ immediately above or below  $r_{\rm LR}$) $\beta_\pm$ may become  negative. In that case the energy and angular momentum~\eqref{Eq:EnergyAngMomTimelike} of a timelike particle along such a putative circular orbit become imaginary: such region  will \textit{not} contain TCOs (rather it will have spacelike circular orbits).

We will show now that, for either rotation sense, there is always one side in the immediate vicinity of a LR, wherein TCOs are forbidden, whose relative location with respect to the LR  depends solely on the  stability of the latter.

Assume a LR exists\footnote{One can consider either a prograde or retrograde LR or both.} at $r = r_\pm^\text{LR}$ such that $\beta_\pm(r_\pm^\text{LR}) = 0$. The first order Taylor expansion of $\beta_\pm$ around the LR reads
\begin{equation}
	\beta_\pm(r) =  \beta'_\pm(r_\pm^\text{LR}) \delta r + \mathcal{O}(\delta r^2) \ ,
\end{equation}
where $\delta r \equiv r - r_\pm^\text{LR}$. Thus, the sign of $\beta_\pm$ in the vicinity  of the LR is determined by $\delta r$ and
\begin{equation}\label{Eq:BetaDerivative}
	\beta'_\pm\big|_{\rm LR} =  - 2\left[ \Omega'_\pm \left( g_{t\varphi} + \Omega_\pm g_{\varphi \varphi} \right)\right]_{\rm LR} \ ,
\end{equation}
where we have made use of~\eqref{Eq:LRsEquation} (or equivalently~\eqref{Eq:AngVelTimeLike}).  Explicitly computing $\Omega'_\pm$ (from the  quadratic equation leading to~\eqref{Eq:AngVelTimeLike}),
\begin{equation}
	\Omega'_\pm\big|_{\rm LR} = -\frac{1}{2} \left[\frac{g_{tt}'' + 2 g_{t\varphi}'' \Omega_\pm + g_{\varphi\varphi}'' \Omega_\pm^ 2}{g_{t\varphi}' + \Omega_\pm g_{\varphi\varphi}'}\right]_{\rm LR}\stackrel{\eqref{Eq:StabilityLRs}}{=}  -\frac{1}{2} \frac{V_0''(r_\pm^{\rm LR})}{ L^2_\pm} \left[\frac{g_{t\varphi}^2 - g_{tt} g_{\varphi\varphi}}{g_{t\varphi}' + \Omega_\pm g_{\varphi\varphi}'}\right]_{\rm LR} \ .
\end{equation}
Thus,  we can rewrite Eq. (\ref{Eq:BetaDerivative}) as
\begin{equation}
	\beta'_\pm(r_\pm^\text{LR}) = \frac{V_0''(r_\pm^\text{LR})}{L_\pm^2} \left[\frac{g_{t\varphi} + \Omega_\pm g_{\varphi\varphi}}{g_{t\varphi}' + \Omega_\pm g_{\varphi\varphi}'} \left( g_{t\varphi}^2 - g_{tt} g_{\varphi\varphi} \right)\right]_{\rm LR} \ .
\end{equation}
Using Eqs. \eqref{LRcon1} and \eqref{Eq:LRsEquation} to simplify this result, 
the first order Taylor expansion of $\beta_\pm$ can be finally written as,
\begin{equation}
	\beta_\pm(r) = \frac{V_0''(r_\pm^\text{LR})}{L_\pm^2} \left[\frac{\left( g_{t\varphi}^2 - g_{tt} g_{\varphi\varphi} \right)^3}{(g_{t\varphi}')^2 - g_{tt}' g_{\varphi\varphi}'}\right]^{1/2}_{\rm LR} \delta r + \mathcal{O}(\delta r^2) \ .
\end{equation}
Thus, the sign of $\beta_\pm$ is determined by the signs of $V_0''(r_\pm^\text{LR})$ ( stability of the LR) and $\delta r$ (upper or lower neighbourhood of the LR). It follows that in the  vicinity of:
\begin{itemize}
	\item An  unstable LR ($V_0''(r_\pm^\text{LR}) > 0$), $\beta_\pm(r) < 0$ in the region below the LR, \textit{i.e.}, $r < r_\pm^\text{LR} \Leftrightarrow \delta r < 0$, wherein no TCOs are thus possible. No obstruction exists for TCOs on the other side.

	\item A  stable LR ($V_0''(r_\pm^\text{LR}) < 0$), a symmetric reasoning holds. Thus, no TCOs are  possible in the region above the LR, \textit{i.e.}, $r > r_\pm^\text{LR} \Leftrightarrow \delta r > 0$.

\end{itemize}

\subsection{ Stability}

 It is possible to extend further this analysis and determine the  stability of the TCOs that occur in the neighbourhood of the LR. We will now show that the region above (below) an  unstable (stable) LR always harbours  unstable (stable) circular orbits.

To consider the  stability of TCOs we examine $V_{-1}''(r)$. Using the definitions of energy and angular momentum, Eq. (\ref{Eq:EnergyAngMomTimelike}), we can write,
\begin{equation}\label{Eq:StabilityTimelikeLRs}
	V_{-1}''(r) = \frac{g_{tt}''(g_{t\varphi} + \Omega_\pm g_{\varphi\varphi})^2 - 2 g_{t\varphi}''(g_{tt} + \Omega_\pm g_{t\varphi})(g_{t\varphi} + \Omega_\pm g_{\varphi\varphi}) + g_{\varphi\varphi}''(g_{tt} + \Omega_\pm g_{t\varphi})^2}{\beta_\pm (g_{t\varphi}^2 - g_{tt} g_{\varphi\varphi})} - \frac{(g_{t\varphi}^2 - g_{tt} g_{\varphi\varphi})''}{g_{t\varphi}^2 - g_{tt} g_{\varphi\varphi}} \ .
\end{equation}
 $V_{-1}''(r_\pm^\text{LR}) $ diverges,  since $\beta_\pm(r_\pm^\text{LR}) \rightarrow 0$ features in the denominator of the  first term.  Thus we need to understand with which sign it diverges (and we can ignore the finite second term).

Approaching the LR from the side wherein TCOs are allowed ($\beta_\pm > 0$), the denominator of the first term in Eq. (\ref{Eq:StabilityTimelikeLRs}), is positive (recall~\eqref{Bdef}). Hence, $\beta_\pm (g_{t\varphi}^2 - g_{tt} g_{\varphi\varphi}) > 0$ and the sign of the term is dictated by the numerator.

Considering now the numerator of the first term in eq. (\ref{Eq:StabilityTimelikeLRs}),  Using similar manipulations as before and using eq. (\ref{Eq:StabilityLRs}), the numerator can be written, at $r = r_\pm^\text{LR}$, as,
\begin{equation}
	g_{tt}''(g_{t\varphi} + \Omega_\pm g_{\varphi\varphi})^2 - 2 g_{t\varphi}''(g_{tt} + \Omega_\pm g_{t\varphi})(g_{t\varphi} + \Omega_\pm g_{\varphi\varphi}) + g_{\varphi\varphi}''(g_{tt} + \Omega_\pm g_{t\varphi})^2 = V_0'' (r_\pm^\text{LR}) \frac{(g_{t\varphi}^2 - g_{tt} g_{t\varphi})^2}{L_\pm^2} \ .
\end{equation}
Thus, the sign of the numerator is dictated by $V_0''(r_\pm^\text{LR})$. We concluded that, when approaching the LR from the allowed region:  $V_{-1}''(r_\pm^\text{LR})\rightarrow +\infty$  if the LR is  unstable ($V_0''(r_\pm^\text{LR}) > 0$), and $V_{-1}''(r_\pm^\text{LR})\rightarrow -\infty$ if the LR is  stable ($V_0''(r_\pm^\text{LR}) < 0$). In short:
\begin{itemize}
	\item Near an  unstable LR, $V_0''(r_\pm^\text{LR}) > 0$, the allowed region for TCOs harbours  unstable orbits -  Fig.~\ref{Fig:CompIllustStableLR} (top panel).

	\item Near a  stable LR, $V_0''(r_\pm^\text{LR}) < 0$, the allowed region for TCOs harbours  stable orbits - Fig.~\ref{Fig:CompIllustStableLR} (bottom panel).

	\begin{figure}[h!]
		\centering
\begin{tikzpicture}
			\draw[ultra thick,->] (-0.05,-0.05) -- (10.45,-0.05) node[anchor=north west]{$r$};
			\filldraw[red!10, draw=red!85!black,ultra thick] (0,0) rectangle (5,2) node[pos=0.5,red!85!black,font=\large\bfseries]{TCOs forbidden};
			\draw[dash pattern=on 3pt off 4pt,ultra thick] (5.1,0) -- (5.1,2);
			\filldraw[yellow!10, draw=yellow!20!orange,ultra thick] (5.2,0) rectangle (10.2,2) node[pos=0.5,yellow!20!orange,align=center,font=\large\bfseries]{\textit{Radially}\\ Unstable TCOs};
			\node[font=\bfseries,align=center] at (5.1,2.45){\textit{Radially}\\ Unstable LR};
			\node at (5.1,-0.35){$r_\pm^{\text{LR}}$};
		\end{tikzpicture}
		\begin{tikzpicture}
			\draw[ultra thick,->] (-0.05,-0.05) -- (10.45,-0.05) node[anchor=north west]{$r$};
			\filldraw[green!10, draw=green!70!black,ultra thick] (0,0) rectangle (5,2) node[pos=0.5,green!70!black,align=center,font=\large\bfseries]{\textit{Radially}\\ Stable TCOs};;
			\draw[dash pattern=on 3pt off 4pt,ultra thick] (5.1,0) -- (5.1,2);
			\filldraw[red!10, draw=red!85!black,ultra thick] (5.2,0) rectangle (10.2,2) node[pos=0.5,red!85!black,font=\large\bfseries]{TCOs forbidden};
			\node[font=\bfseries,align=center] at (5.1,2.45){\textit{Radially}\\ Stable LR};
			\node at (5.1,-0.35){$r_\pm^{\text{LR}}$};
		\end{tikzpicture}
		\caption{Structure of the equatorial TCOs in the vicinity an  unstable (top panel) and stable (bottom panel) LR.}
		\label{Fig:CompIllustStableLR}
	\end{figure}
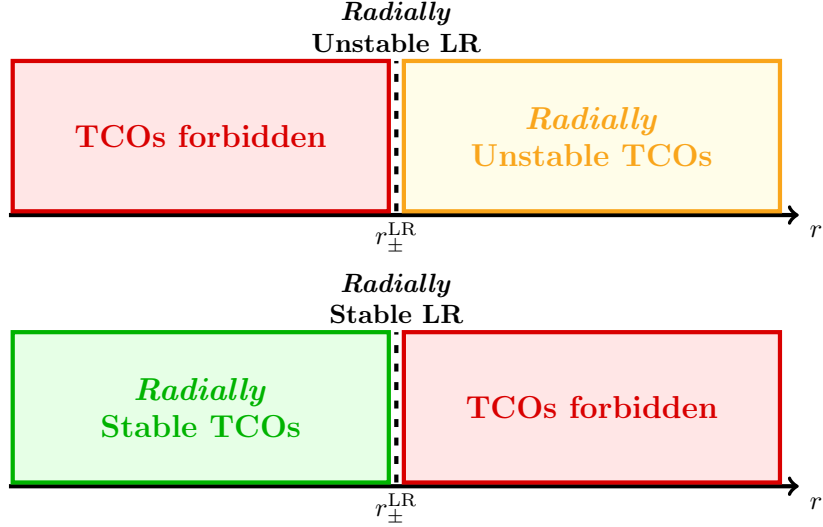
	
\end{itemize}

\subsection{Generality}
The analysis above has two interesting corollaries.

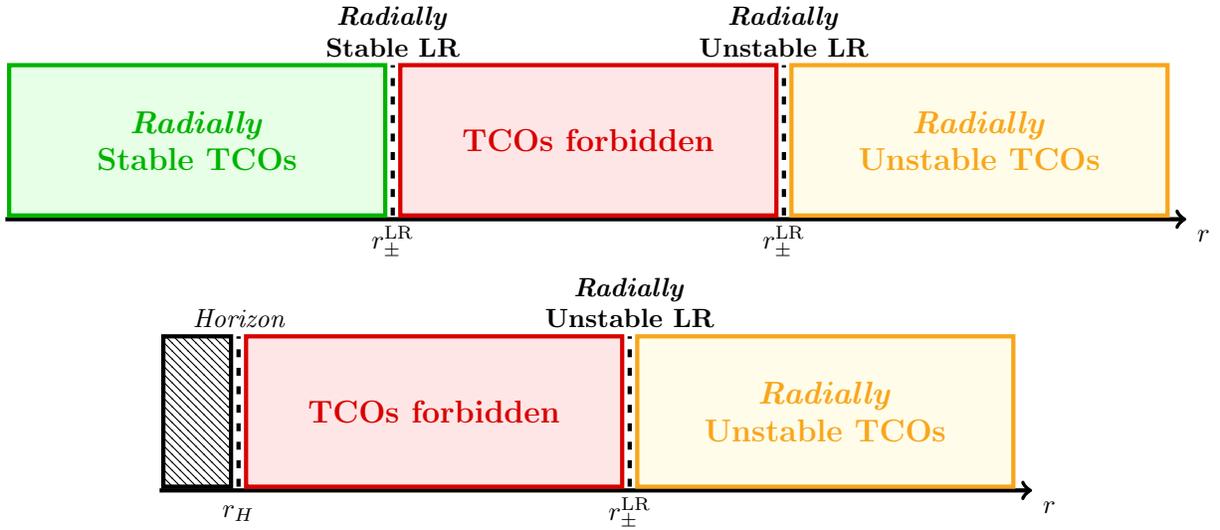
\begin{figure}[h!]
	\centering
\begin{tikzpicture}
		\draw[ultra thick,->] (-0.05,-0.05) -- (15.65,-0.05) node[anchor=north west]{$r$};
		\filldraw[green!10, draw=green!70!black,ultra thick] (0,0) rectangle (5,2) node[pos=0.5,green!70!black,align=center,font=\large\bfseries]{\textit{Radially}\\ Stable TCOs};
		\draw[dash pattern=on 3pt off 4pt,ultra thick] (5.1,0) -- (5.1,2);
		\filldraw[red!10, draw=red!85!black,ultra thick] (5.2,0) rectangle (10.2,2) node[pos=0.5,red!85!black,font=\large\bfseries]{TCOs forbidden};
		\draw[dash pattern=on 3pt off 4pt,ultra thick] (10.3,0) -- (10.3,2);
		\filldraw[yellow!10, draw=yellow!20!orange,ultra thick] (10.4,0) rectangle (15.4,2) node[pos=0.5,yellow!20!orange,align=center,font=\large\bfseries]{\textit{Radially}\\ Unstable TCOs};
		\node[font=\bfseries,align=center] at (5.1,2.45){\textit{Radially}\\ Stable LR};
		\node[font=\bfseries,align=center] at (10.3,2.45){\textit{Radially}\\ Unstable LR};
		\node at (5.1,-0.35){$r_\pm^{\text{LR}}$};
		\node at (10.3,-0.35){$r_\pm^{\text{LR}}$};
	\end{tikzpicture}
	\begin{tikzpicture}
		\filldraw[black!10, pattern=north west lines, draw=black,ultra thick] (0,0) rectangle (0.9,2);
		\draw[ultra thick,->] (-0.05,-0.05) -- (11.55,-0.05) node[anchor=north west]{$r$};
		\draw[dash pattern=on 3pt off 4pt,ultra thick] (1,0) -- (1,2);
		\filldraw[red!10, draw=red!85!black,ultra thick] (1.1,0) rectangle (6.1,2) node[pos=0.5,red!85!black,font=\large\bfseries]{TCOs forbidden};
		\draw[dash pattern=on 3pt off 4pt,ultra thick] (6.2,0) -- (6.2,2);
		\filldraw[yellow!10, draw=yellow!20!orange,ultra thick] (6.3,0) rectangle (11.3,2) node[pos=0.5,yellow!20!orange,align=center,font=\large\bfseries]{\textit{Radially}\\ Unstable TCOs};
		\node[font=\bfseries] at (1,2.25){{\it Horizon}};
		\node[font=\bfseries,align=center] at (6.2,2.45){\textit{Radially}\\ Unstable LR};
		\node at (1,-0.35){$r_H$};
		\node at (6.2,-0.35){$r_\pm^{\text{LR}}$};
	\end{tikzpicture}
	\caption{Structure of the equatorial TCOs for a stationary, axisymmetric, asymptotically flat and $\mathbb{Z}_2$ symmetric: (top panel) horizonless ultra-compact object around its pair of LRs; (bottom panel) BH around its unstable LR.  These illustrations are universal, regardless of the direction of rotation of the LR and of the timelike particle.}
	\label{Fig:CompIllustBHs}
\end{figure}

First, it was  shown  in \cite{Cunha:2017qtt}  that for asymptotically flat stationary and axisymmetric horizonless ultracompact objects, that can be smoothly deformed  into flat spacetime, LRs come in  pairs with one stable and one unstable LR. The proof presented above shows that, for such objects with a $\mathbb{Z}_2$ symmetry, the region between the LRs has no TCOs. Otherwise, there would be a subregion between the LRs wherein $\beta_\pm > 0$, which would imply, by continuity, two points with $\beta_\pm = 0$, $i.e.$ another pair of LRs - Fig.~\ref{Fig:CompIllustBHs} (top panel).

A second corollary applies to BHs. It has been shown that a stationary, axisymmetric and asymptotically flat black hole always has (at least) one unstable LR in each sense of rotation \cite{Cunha:2020azh}. This statement, together with our proof, imply that the region between the event horizon and the unstable LR is always a region without  TCOs. This follows from a similar argument to that presented before for horizonless objects. As  in the previous paragraph, this can be easily established by contradiction, and relying on the continuity of $\beta_\pm$ - Fig.~\ref{Fig:CompIllustBHs} (bottom panel).

\section{Illustrations and efficiency}\label{Sec:IllustrationsEfficiency}

As discussed in  the Introduction, the efficiency of a given compact object can be understood as the amount of gravitational energy which is converted into radiation as a timelike particle falls down from infinity. If one assumes that all radiation escapes towards infinity, then the efficiency is computed as the difference between the energy per unit mass measured at infinity and at the ISCO, as given by \eqref{Eq:Efficiency}. This definition of efficiency is only an approximation; the real efficiency should take into account how much of the converted radiation effectively reaches infinity and how much falls back into the BH. However, this simple estimate provides an intuition about the magnitude of the process. Moreover, it provides a simple estimate to compare different models to the Kerr BH, which has a maximal efficiency of $42\%$, probing if alternative models of compact objects could produce even larger energy conversions.

To compute the efficiency, $cf.$ Eq. (\ref{Eq:Efficiency}), we need to compute the energy of the TCO at the ISCO. For Kerr BHs, the location of the ISCO is unambiguous: there is only one solution of~\eqref{isco1} (for each  rotation direction) and no solution of~\eqref{isco21}, thus the ISCO is the same as the MSCO. However, for more generic models, there can be several disconnected regions with  stable TCOs - see $e.g.$~\cite{Delgado:2020udb,Delgado:2020hwr,Collodel:2021gxu}. 
For those objects, the ISCO is no longer the same as the MSCO. 

 The rationale that the efficiency is related to the energy conversion by particles moving along a continuous sequence of stable TCOs, from large distances until the last stable TCO,
suggests the efficiency should be computed on that last stable TCO
that is continuously connected by a sequence of stable TCOs to infinity. This corresponds to MSCO and the corresponding  efficiency is denoted by $\epsilon_{\rm MSCO}$. 
In parallel, we shall also consider an alternative efficiency computed at the ISCO, denoting the corresponding  efficiency by $\epsilon_{\rm ISCO}$. As we shall see 
\begin{equation}
 \epsilon_{\rm ISCO}\geqslant  \epsilon_{\rm MSCO} \ .
\end{equation}

In the following we will analyse the efficiency of several stationary, axisymmetric and asymptotically flat spinning horizonless compact objects (that we shall generically refer to as `stars') as well as of spinning BHs. The examples include: mini-boson stars~\cite{SCHUNCK1998389,PhysRevD.56.762}, gauged  boson stars~\cite{Delgado:2016jxq}, axionic boson stars~\cite{Guerra:2019srj,Delgado:2020udb}, Proca stars~\cite{Brito:2015pxa,Herdeiro:2017fhv,Minamitsuji:2018kof}, Kerr BHs with synchronised axionic hair~\cite{Delgado:2020hwr} and BHs of the shift-symmetric Horndeski theory (Einstein-scalar-Gauss-Bonnet BHs)~\cite{Sotiriou:2014pfa,Delgado:2020rev}.  
The selected configurations are representative in each case. Also, they correspond to alternative models in and beyond GR theories that are well motivated, with extensive literature discussing various features of the solutions. A non-exhaustive list of references, beyond the original papers cited above, is given by the following works \cite{Berti:2015itd,Cao:2016zbh,Vincent:2015xta,Cunha:2016bjh,Ni:2016rhz,Cunha:2015yba,Herdeiro:2021lwl,Herdeiro:2020kba,Shen:2016acv,Bryant:2021xdh}.

For completeness, let us briefly comment on how the
considered configurations have been found.
For  both stars and BHs, 
the same methodology has been  used to obtain the  solutions.
In each case, one starts with the action of the theory and obtained first the equations of motion.
 That is, after defining an appropriate ansatz,
 we have computed the set of partial differential equations 
 for the metric (and matter fields), together with the corresponding boundary conditions. 
 Unfortunately, in all cases, no analytical solutions  are known to exist.
Therefore, for all models, the solutions were found by 
employing a professional numerical solver \cite{solver1,solver3}, which uses a Newton-Raphson method.
A detailed aspects of these aspects can be found in the papers
where the solutions were initially reported,
see $e.g.$ Ref.  \cite{Herdeiro:2015gia}.
Once the (numerical) components of the metric are known, 
one can do various physical and phenomenological studies.

With the knowledge learned from the previous Sections, 
we have considered the stability 
of TCOs for all solutions in all  mentioned examples, 
and investigated in which regions of the spacetime 
it was possible to have stable TCOs ($V''_{-1} < 0$), unstable TCOs ($V''_{-1} > 0$), no TCOs ($\beta_\pm < 0$), or no circular orbits at all ($\Omega_\pm \in \mathbb{C}$). For solutions that possess LRs, we also computed their radii, by solving Eq. \eqref{Eq:LRsEquation} together with Eq. \eqref{sigmapm}, as well as their stability, Eq. \eqref{Eq:StabilityLRs}. With all regions defined, 
we have analysed their boundaries, mainly the boundaries between regions of stable and unstable TCOs, as well as, between regions without any circular orbits and stable TCOs. 
At each boundary of interest, we 
have computed the energy of such circular orbit together with the efficiency.

\subsection{Stars}

All star solutions discussed herein are only known numerically  (no analytic form is known, although in some cases perturbative expansions are possible, $e.g.$~\cite{Delgado:2020rev}). The solutions are computed specifying an ansatz for the metric and remaining fields. For the problem at hand, however, we only need the metric. Thus, we shall only specify the ansatz metric, which is the same for all stars considered, and reads
\begin{equation}
	ds^2 = -e^{2F_0} dt^2 + e^{2F_1} \left( dr^2 + r^2 d\theta^2 \right) + e^{2F_2} r^2 \sin^2 \theta \left( d\varphi - \frac{W}{r} dt \right)^2~,
\end{equation}
where $F_0,F_1,F_2$ and $W$ are ansatz functions that depend solely on the radial and co-latitude coordinates $(r,\theta)$.
 The correspondence with the ansatz~\eqref{metric1} is:
\begin{equation}
g_{tt}=-e^{2F_0}+e^{2F_2}  W^2 \sin^2\theta \ , \ g_{rr}=e^{2F_1}  \ , \ g_{\theta \theta}=e^{2F_1} r^2 \ , \  g_{\varphi \varphi}=e^{2F_2} r^2 \sin^2\theta \ ,  \ g_{t\varphi}=-e^{2F_2} r W \sin^2\theta  \ .
\end{equation}

For each family of star solutions we shall present four plots. On the one hand, the top (bottom) two plots exhibit the results for prograde (retrograde) orbits. 
On the other hand, the left plots illustrate the  structure of TCOs and LRs $vs.$ the radial coordinate $r$ (which is normalise for each family), in the space of solution. For that, the specific solution is labelled by the maximal value of the scalar field $\phi_\text{max}$ (except for the Proca stars). In this way, each horizontal line corresponds exactly to one star solution.  Then, for each plot there are four
different coloured regions: in violet, we have a region in which no (timelike, null or spacelike) circular orbits exits (labelled \textit{No COs}); in red, we have a region in which no TCOs exist (labelled \textit{No TCOs}); In yellow, the region of  unstable TCOs (labelled  \textit{UTCOs}); in green, the region of  stable TCOs (labelled \textit{STCOs}). The plots also exhibit the MSCO,  ISCO, LRs and a solid black horizontal line representing the first solution for which MSCO $\neq$ ISCO.
In all cases, the results were found by  extrapolation  into the continuum the data 
corresponding to a large number (from a  few hundreds to thousands) of individual points.

For all stars studied in this work, the structure of TCOs close to LRs follows exactly the patterns deduced in the previous section. In particular, they only possess a pair of retrograde LRs, in which the LR with the largest (smallest) radii is always  unstable (stable). Then, the region above (below) an unstable (stable) LR is a region of  unstable (stable) TCOs, and the region between the pair of LRs is a region without TCOs. 

Finally, the right plots exhibit the efficiencies $\epsilon_{\rm MSCO}$ and $\epsilon_{\rm ISCO}$ $vs.$ $\phi_\text{max}$: $\epsilon_{\rm ISCO}$ is given by the solid red line, whereas $\epsilon_{\rm ISCO}$  is given by dashed green line. For convenience, we keep also the same black solid line as in the left plots.

\subsubsection{Mini-Boson Stars}

Mini-boson stars are regular everywhere solutions of the (complex-)Einstein-Klein-Gordon theory, where a massive free scalar field $\Psi$ is minimally coupled to Einstein's gravity. The action can be written as,
\begin{equation}
	\mathcal{S} = \int d^4 x \sqrt{-g} \left[ \frac{R}{16\pi} - g^{\mu\nu} \partial_\mu \Psi^* \partial_\nu \Psi - \mu^2 \Psi^* \Psi \right] \ ,
\end{equation} 
where $\mu$ is the mass of the scalar field. 
These solutions can be consider as a macroscopic version of a Bose-Einstein condensate and were initially developed (in spherical symmetry) by Kaup\cite{PhysRev.172.1331} and Ruffini and Bonazzala\cite{PhysRev.187.1767} - see also, $e.g.$~\cite{Schunck:2003kk,Herdeiro:2017fhv}. Later, due to the efforts of Schunck and Mielke\cite{SCHUNCK1998389} and Yoshida and Eriguchi\cite{PhysRevD.56.762}, spinning generalisation of the previous static solutions were found - see also, $e.g.$~\cite{Herdeiro:2019mbz}.

Here we will consider three families of mini-boson stars with two different values of the azimuthal harmonic index, $m = \{1, 2\}$, which appears in the scalar field ansatz,
\begin{equation}\label{Eq:AnsatzScalarField}
	\Psi = \phi(r,\theta) e^{i(m \varphi - \omega t)} \ ,
\end{equation}
where $\phi$ is a $\{r,\theta\}$ dependent scalar field amplitude and $\omega$ is the angular frequency of the scalar field.
We focus on the fundamental states only (even parity, nodeless BSs).

For the $m=1$ solutions, the study of the efficiency was already done in \cite{Collodel:2021gxu}. These authors, however, focused  on solutions for which MSCO = ISCO. Here we will also consider solutions for which MSCO $\neq$ ISCO and compute $\epsilon_{\rm MSCO} \neq \epsilon_{\rm ISCO}$.

\begin{figure}
	\centering
	\includegraphics[scale=0.4]{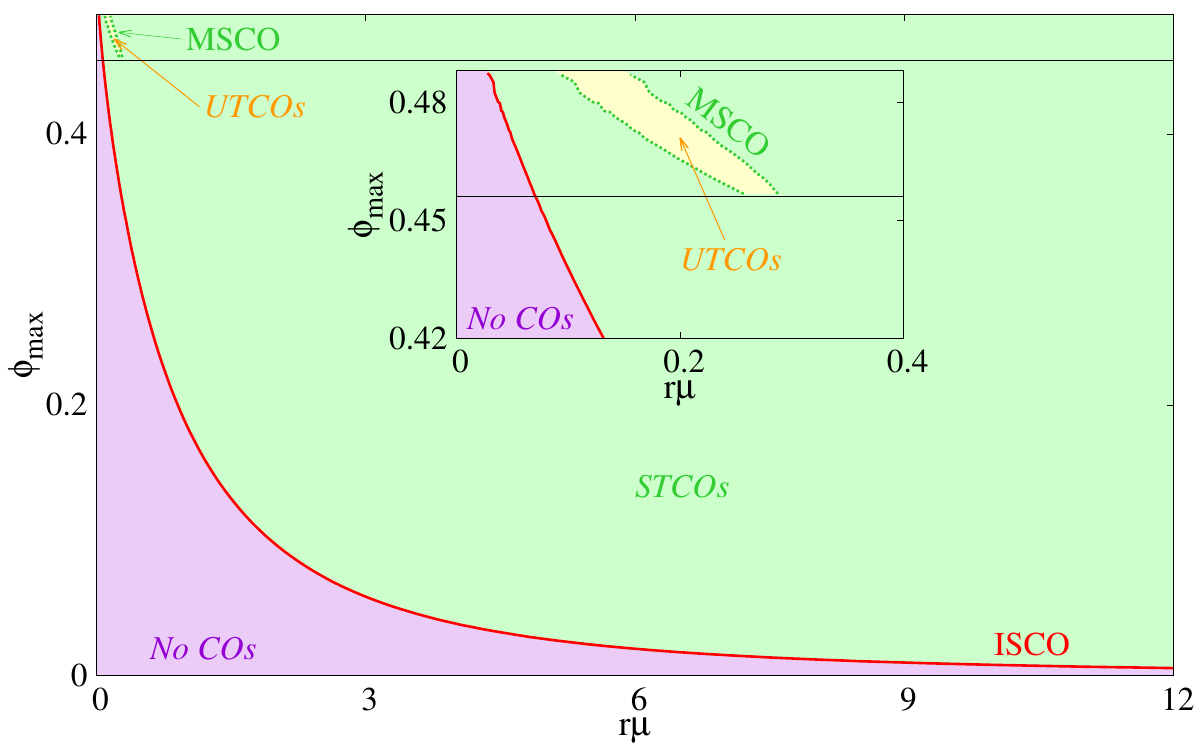}
	\includegraphics[scale=0.4]{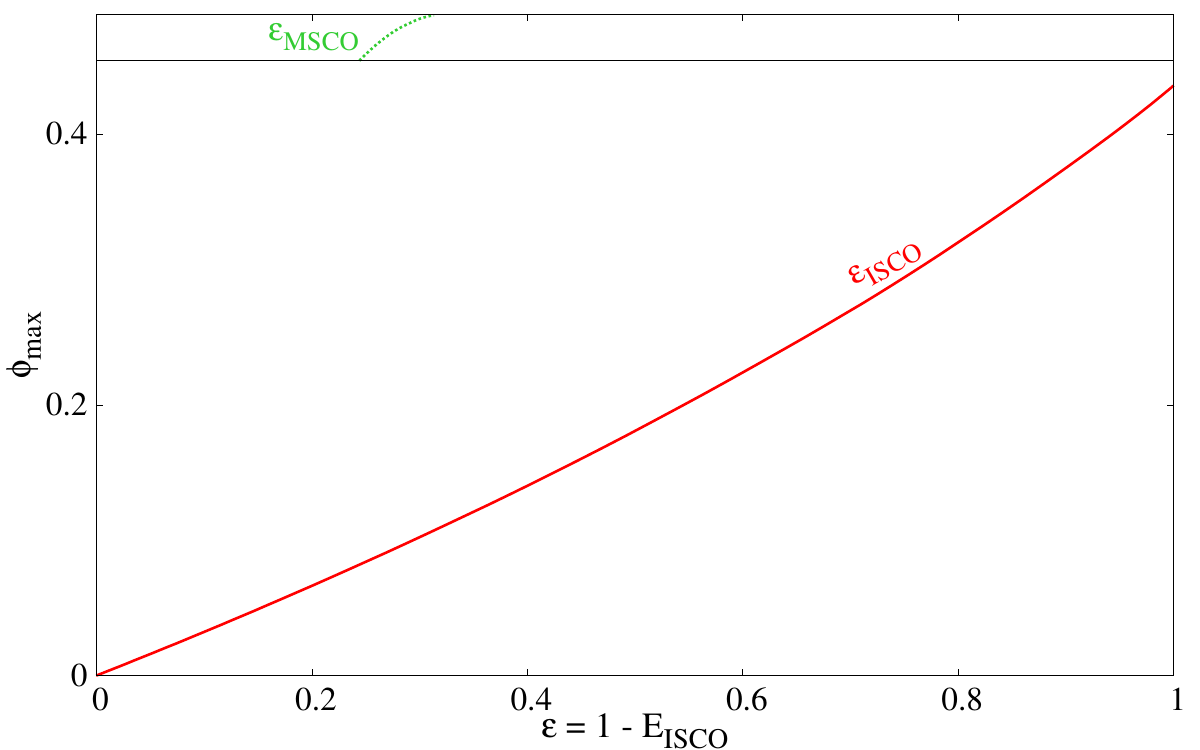}\\
	\includegraphics[scale=0.4]{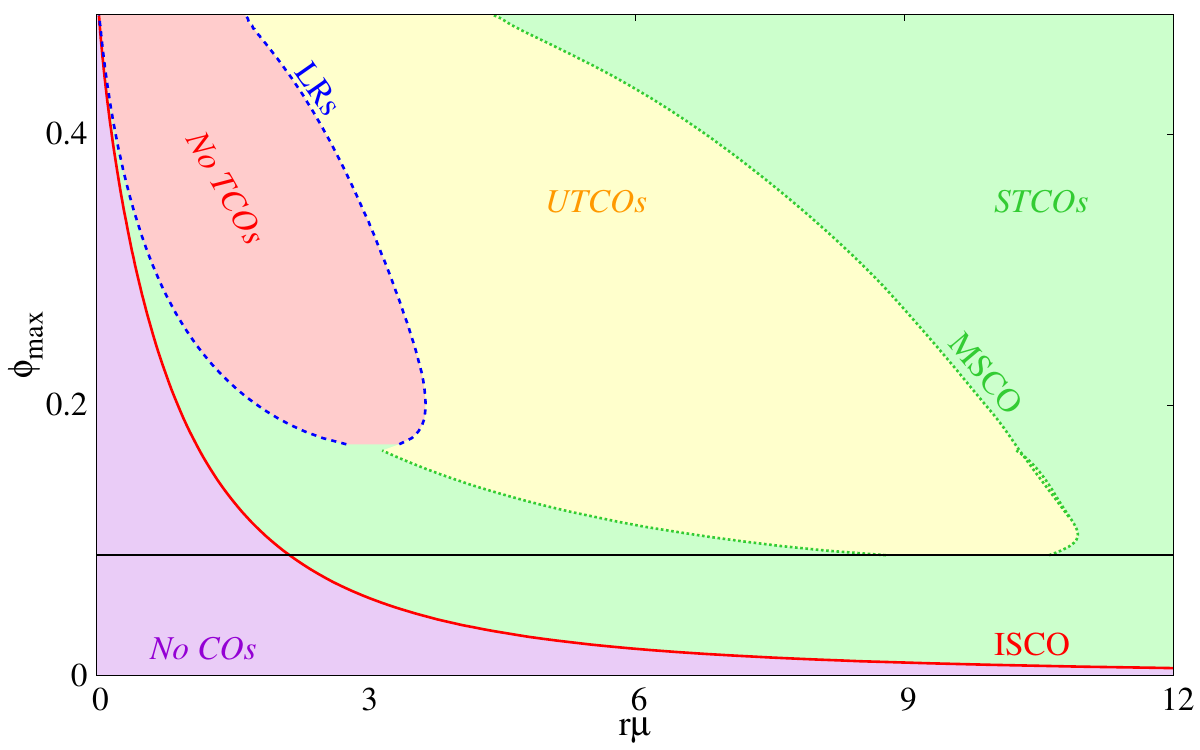}
	\includegraphics[scale=0.4]{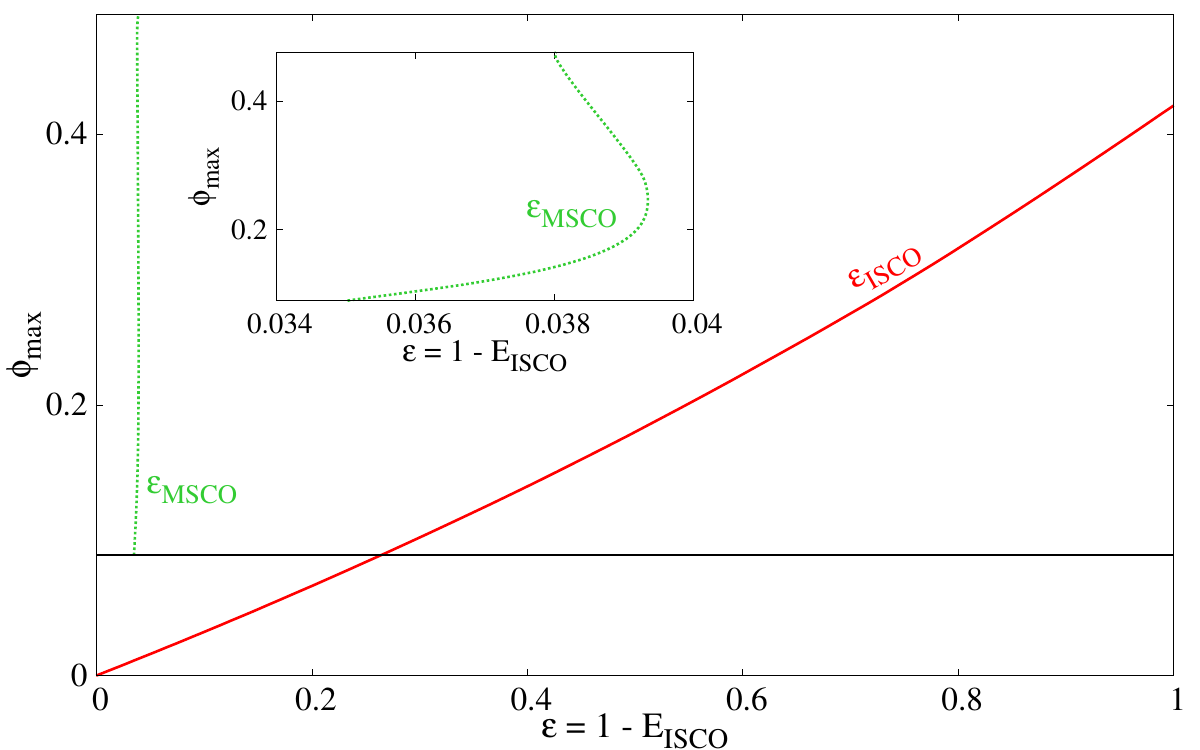}
	\caption{Structure of TCOs and LRs (\textit{left column}) and efficiency (\textit{right column}) for mini-boson stars with $m = 1$. prograde orbits are presented in the \textit{top row}; retrograde orbits are presented in the \textit{bottom row}.}
	\label{Fig:MiniBS_m_1}
\end{figure}

The right panels in Fig. \ref{Fig:MiniBS_m_1} exhibit the efficiency for prograde (top) and retrograde (bottom) TCOs for mini-boson stars with $m=1$. 
In both cases, $\epsilon_{\rm ISCO}$ (red solid line) increases monotonically with $\phi_{\rm max}$ reaching unity. 
In fact, for very compact solutions, the gravitational potential can be deep enough to yield efficiencies greater than one;  truncating  these plots (and the upcoming ones) at $\epsilon = 1$ is, however, enough to show that larger efficiencies than the ones found for Kerr are attained.
Such behaviour is explained by the increasingly smaller radii for the ISCO, which, in turn, leads to progressively smaller energies for TCOs therein. This is consistent with the results in \cite{Collodel:2021gxu} (in the region analysed therein). 

Now consider $\epsilon_{\rm MSCO}$. In the case of prograde orbits, MSCO $\neq$ ISCO only in the strong gravity regime, wherein the solutions start to develop a small region of  unstable TCOs. Then, $\epsilon_{\rm MSCO}$ ranges from $\sim 25\%$ up to $\sim 30\%$.
For retrograde orbits, MSCO $\neq$ ISCO for lower values of  $\phi_{\rm max}$ than in the prograde case.  Thus, for solution which are not very compact 
{(fairly small value of  $\phi_{\rm max} $)},
one can compute $\epsilon_{\rm MSCO}$ which is about $\sim 4\%$ and stabilises around this value even for more compact solutions. 

\begin{figure}
	\centering
	\includegraphics[scale=0.4]{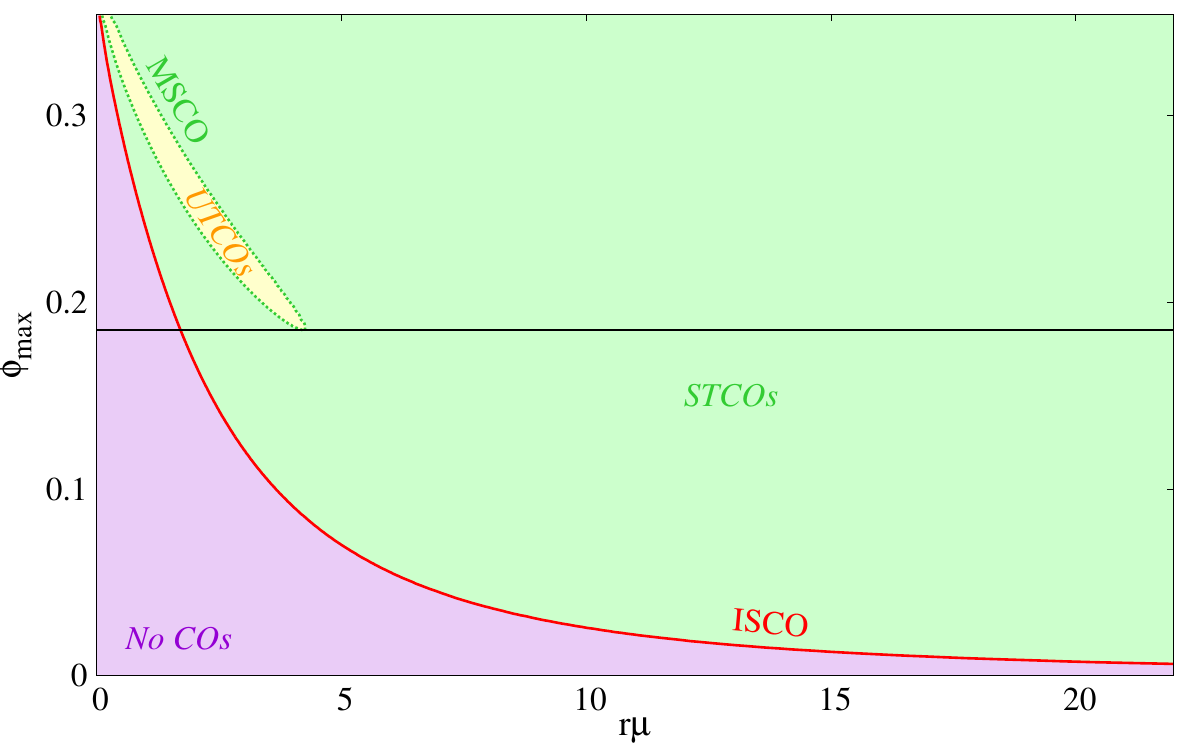}
	\includegraphics[scale=0.4]{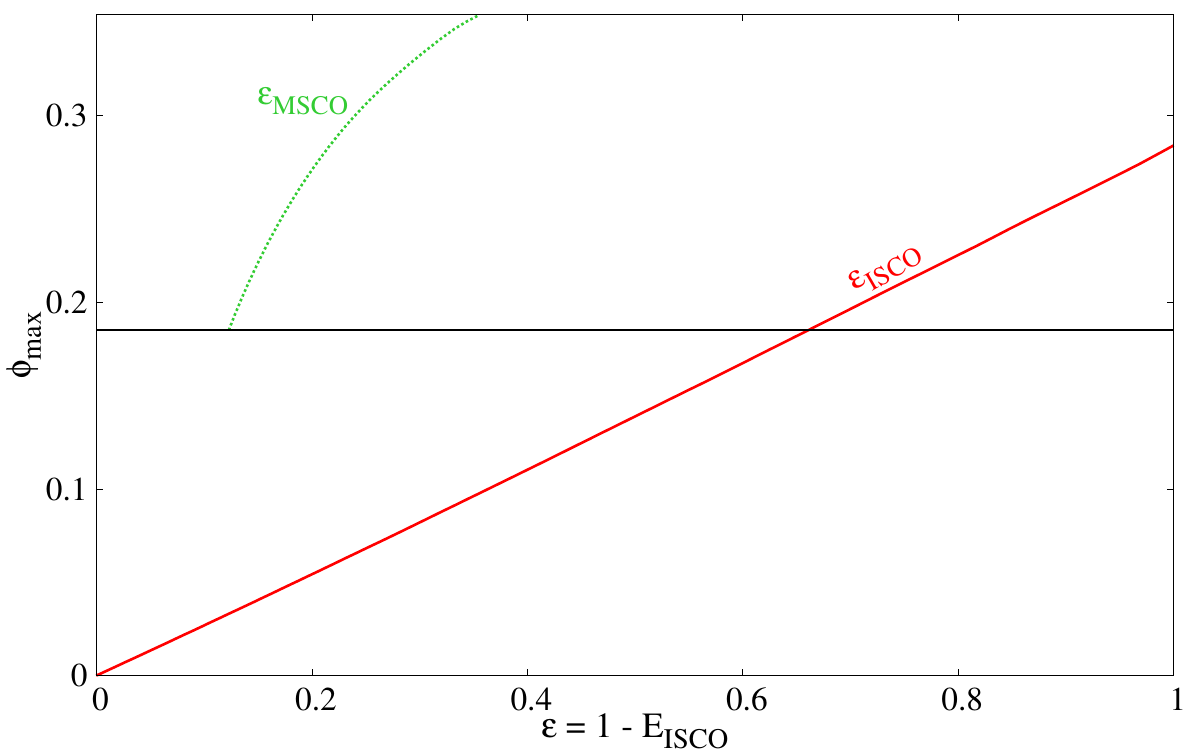}\\
	\includegraphics[scale=0.4]{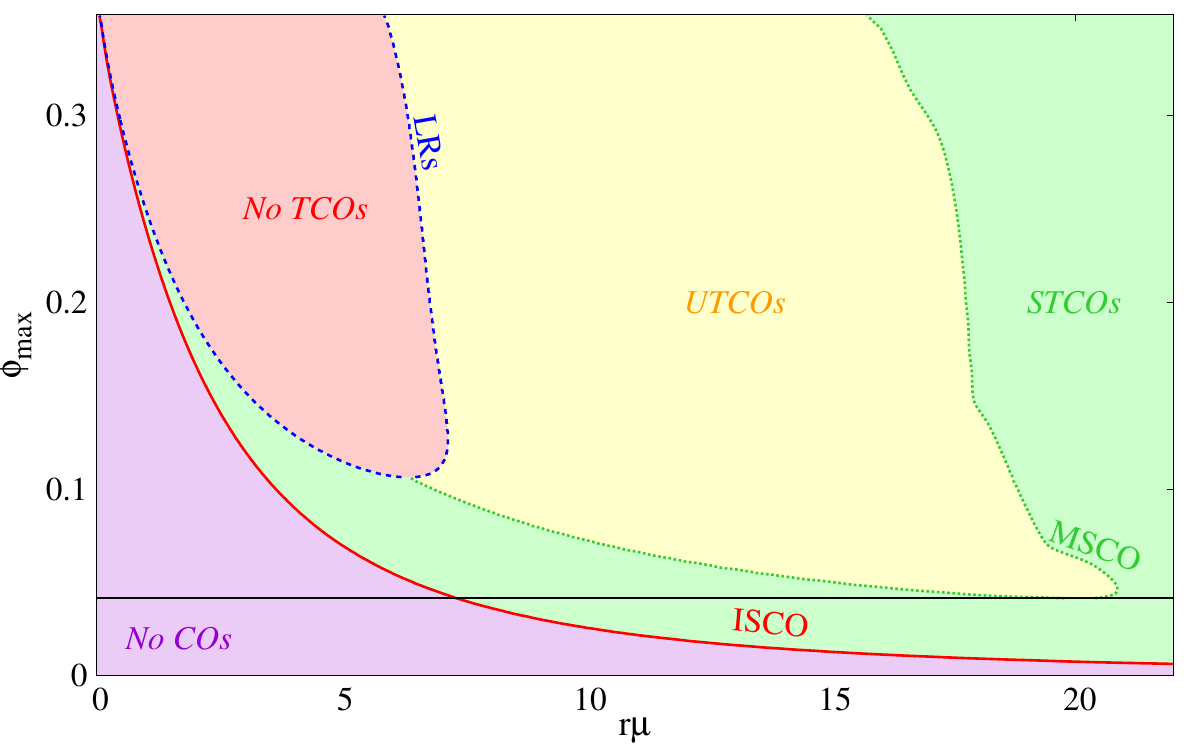}
	\includegraphics[scale=0.4]{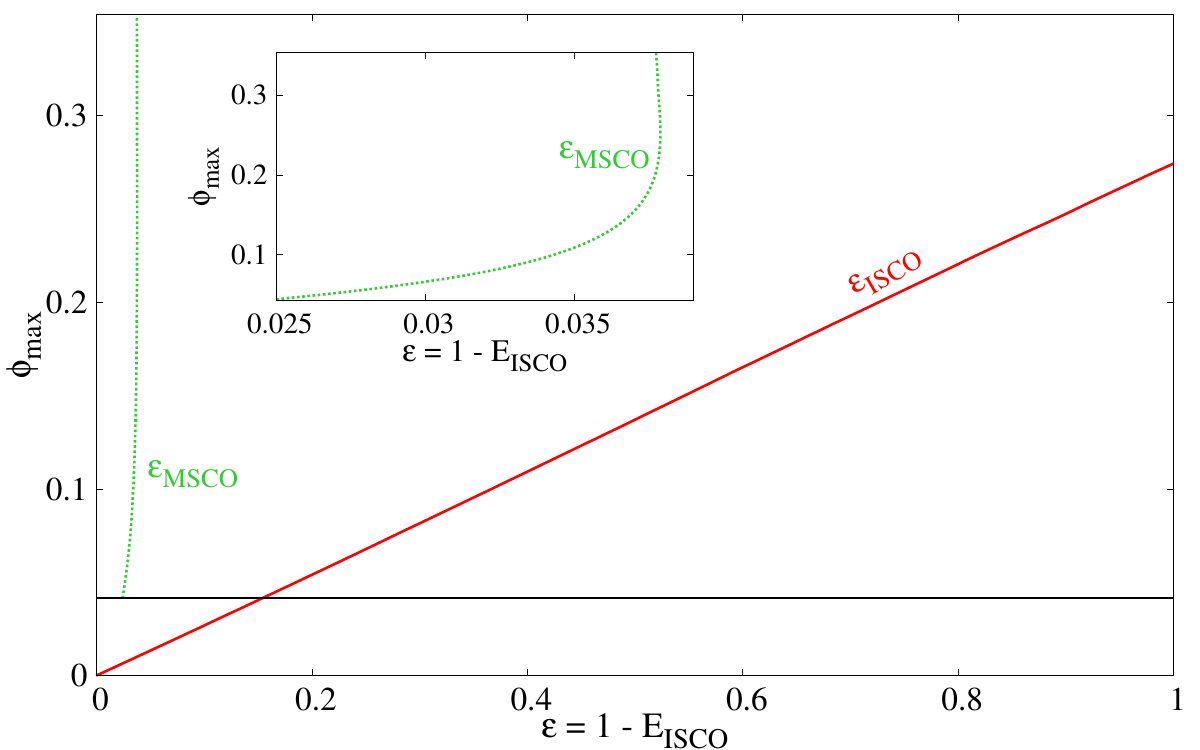}
	\caption{Same as Fig.~\ref{Fig:MiniBS_m_1} but for mini-boson stars with $m = 2$.}
	\label{Fig:MiniBS_m_2}
\end{figure}

In Fig. \ref{Fig:MiniBS_m_2}  the same analysis as in Fig. \ref{Fig:MiniBS_m_1} is repeated for mini-boson stars with $m=2$. The overall structure of TCOs and LRs as well as the efficiencies are very similar as for the $m=1$ case. The most notorious difference is that the region of  unstable TCOs, both for co- and retrograde orbits, occurs for smaller value of $\phi_{\rm max}$. 
For prograde orbits (top), MSCO $\neq$ ISCO above $\phi_{\rm max} \sim 0.18$, wherein,  $\epsilon_{\rm ISCO}\sim 66\%$ and $\epsilon_{\rm MSCO}\sim 12\%$. For more compact solutions, $\epsilon_{\rm MSCO}$ increases monotonically until $\sim 35\%$, whereas $\epsilon_{\rm ISCO}$ reaches unity.
For retrograde orbits (bottom), MSCO $\neq$ ISCO for lower values of $\phi_{\rm max}$. For the first solution with $\epsilon_{\rm MSCO}\neq \epsilon_{\rm ISCO}$, $\epsilon_{\rm ISCO}\sim 15\%$ and  $\epsilon_{\rm MSCO}\sim 2.5\%$. Increasing the compactness, these values increase to around $\epsilon_{\rm MSCO}\sim 3.8\%$ and  $\epsilon_{\rm ISCO}\sim 100\%$.

\subsubsection{Adding scalar field self-interaction: Axion boson stars}
Axion boson stars are solutions of the (complex)-Einstein-Klein-Gordon theory where a massive complex scalar field $\Psi$ with self-interactions is minimally coupled to Einstein's gravity. The action is,
\begin{equation}\label{Eq:ActionAxion}
	\mathcal{S} = \int d^4 x \sqrt{-g} \left[ \frac{R}{16\pi} - g^{\mu\nu} \partial_\mu \Psi^* \partial_\nu \Psi - V(|\Psi|^2) \right]  \ ,
\end{equation}
where $V(|\Psi|^2)$ denotes the self-interactions of the scalar field. Assuming the ansatz (\ref{Eq:AnsatzScalarField}) for the complex scalar field, we can write the self-interaction potential based on the QCD axion potential\cite{diCortona:2015ldu} to which we add a constant in order to have asymptotically flat solutions,
\begin{equation}\label{Eq:AxionPotential}
	V(\phi) = \frac{2\mu_a^2 f_a}{B} \left[ 1 - \sqrt{1 - 4B \sin^2 \left( \frac{\phi}{2 f_a} \right) } \right] \ ,
\end{equation}
where $B$ is a constant defined by the quark masses, $B  \approx 0.22$. The physical meaning of $\mu_a$ and $f_a$ can be easily seen by performing an expansion of the potential around $\phi = 0$,
\begin{equation}
	V(\phi) = \mu_a^2 \phi^2 - \left( \frac{3B - 1}{12} \right) \frac{\mu_a^2}{f_a^2} \phi^4 + \dots \ .
\end{equation}
Thus, $\mu_a$ determines the mass of the axion-like particle, while $f_a$ is related to the self-interaction quartic coupling. We shall refer to $\mu_a$ and $f_a$ as the axion-like particle's mass and decay constant. 
As  $f_a \rightarrow \infty$, mini-boson stars are recovered.

The first study about these stars was presented in \cite{Guerra:2019srj}, for the spherical case. A spinning generalisation was presented later in \cite{Delgado:2020udb}. Here, to probe the effect of the self-interactions on the structure of TCOs and $\epsilon_\pm$, we will consider two small values of the decay constant, $f_a = \{0.03, 0.05\}$. The corresponding results are presented in \cite{Delgado:2020udb}.

\begin{figure}
	\centering
	\includegraphics[scale=0.4]{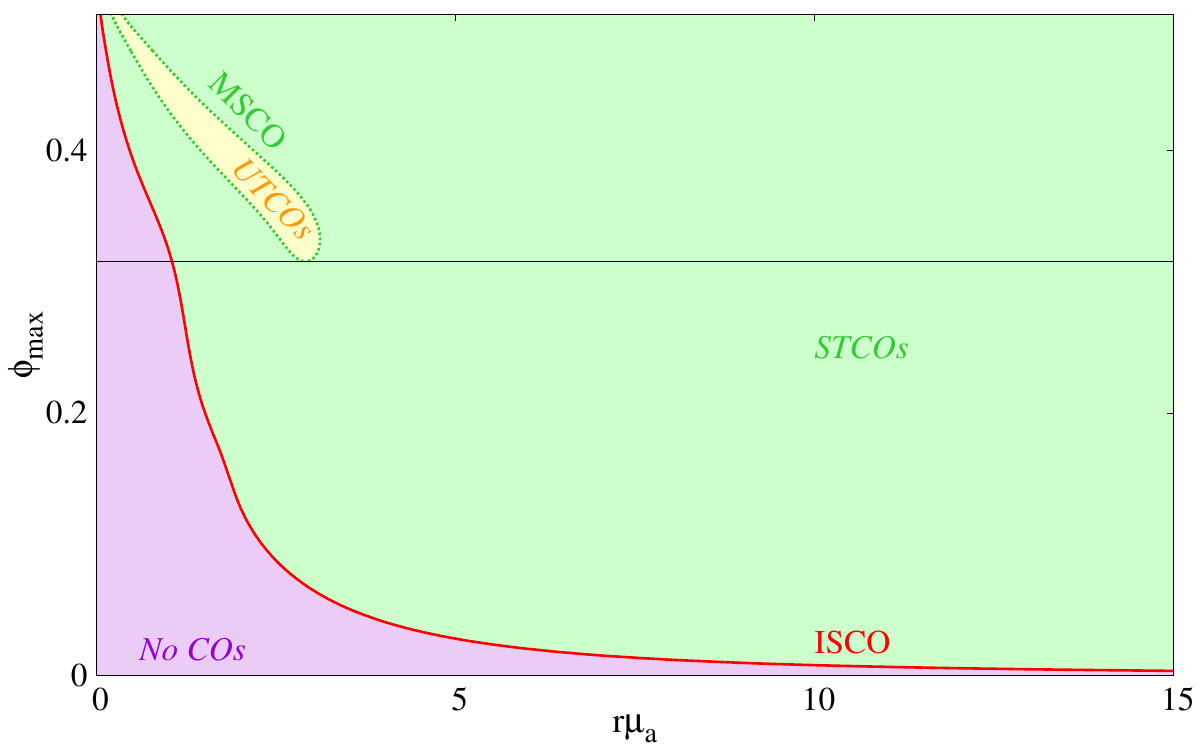}
	\includegraphics[scale=0.4]{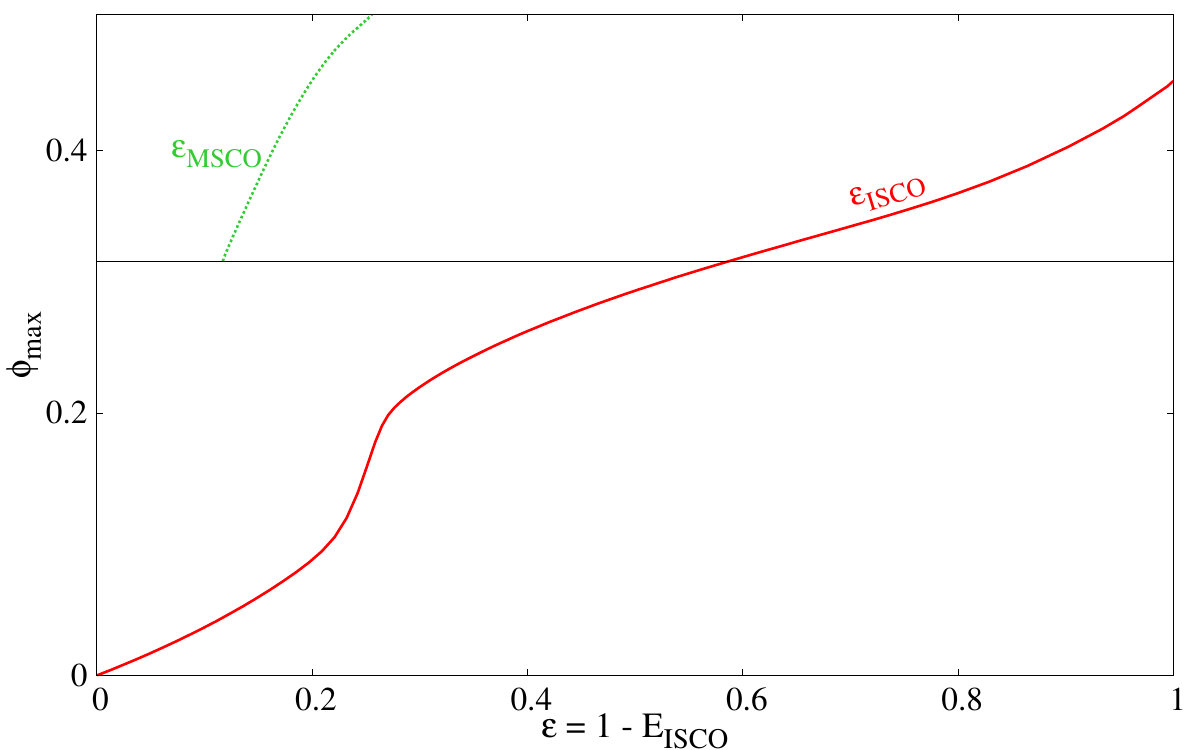}\\
	\includegraphics[scale=0.4]{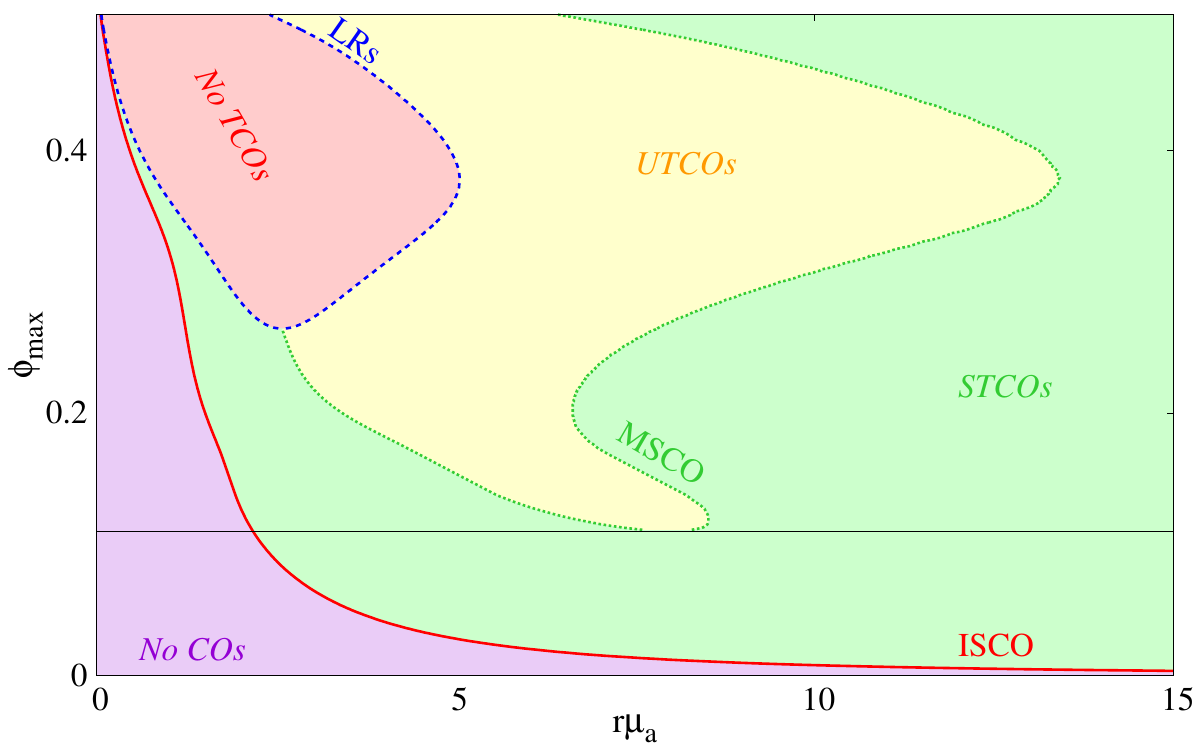}
	\includegraphics[scale=0.4]{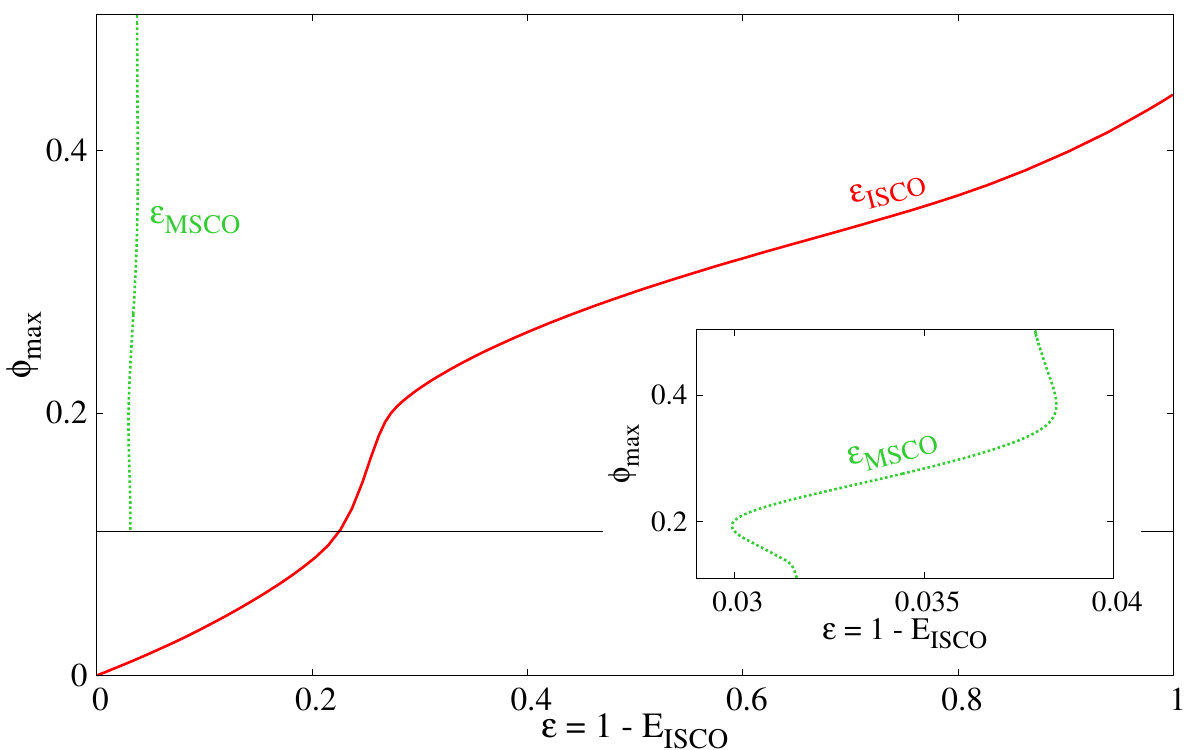}
	\caption{Same as Fig.~\ref{Fig:MiniBS_m_1} but  for axion boson stars with $f_a = 0.05$.} 
	\label{Fig:AxionBS_fa_005}
\end{figure}

In Fig. \ref{Fig:AxionBS_fa_005} we consider $f_a = 0.05$ -- a more detailed analysis of the structure of TCOs is done in \cite{Delgado:2020udb}. We can see that both the structure of TCOs and LRs, as well as the efficiency follow similar patterns to the previous cases. In particular, $\epsilon_{\rm ISCO}$ grows monotonically towards unity, moving towards the strong gravity regime. 
For prograde orbits, at the first solution when  MSCO $\neq$ ISCO, $\epsilon_{\rm ISCO}\sim 59\%$ and $\epsilon_{\rm MSCO}\sim 12\%$. After this solution, $\epsilon_{\rm MSCO}$ increases monotonically until $\sim 26\%$.
For retrograde orbits, at the first solution when  MSCO $\neq$ ISCO, $\epsilon_{\rm ISCO}\sim 23\%$ and  $\epsilon_{\rm MSCO}\sim 3.3\%$. The latter remains approximately constant for other solutions, with a local minimum of $\epsilon_{\rm MSCO}\sim 3\%$ at $\phi_{\rm max} \approx 0.193$ and $\epsilon_{\rm MSCO}\sim 3.8\%$ for the largest values of $\phi_{\rm max}$.

\begin{figure}
	\centering
	\includegraphics[scale=0.4]{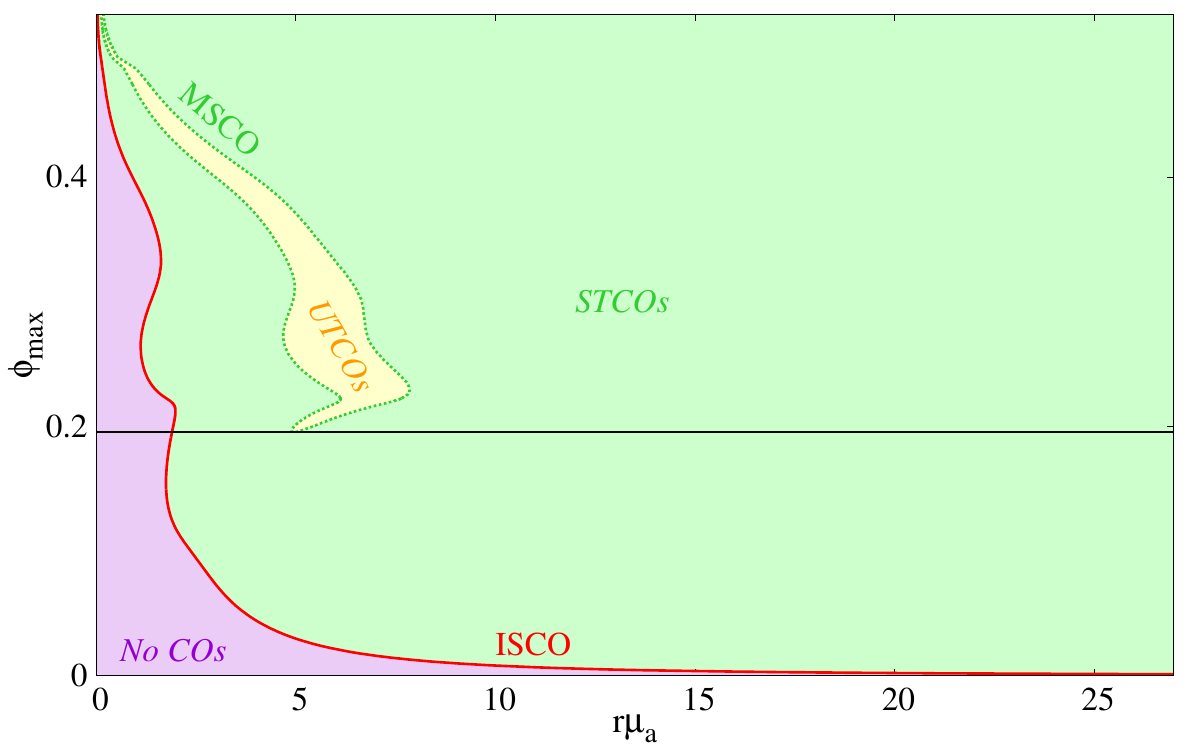}
	\includegraphics[scale=0.4]{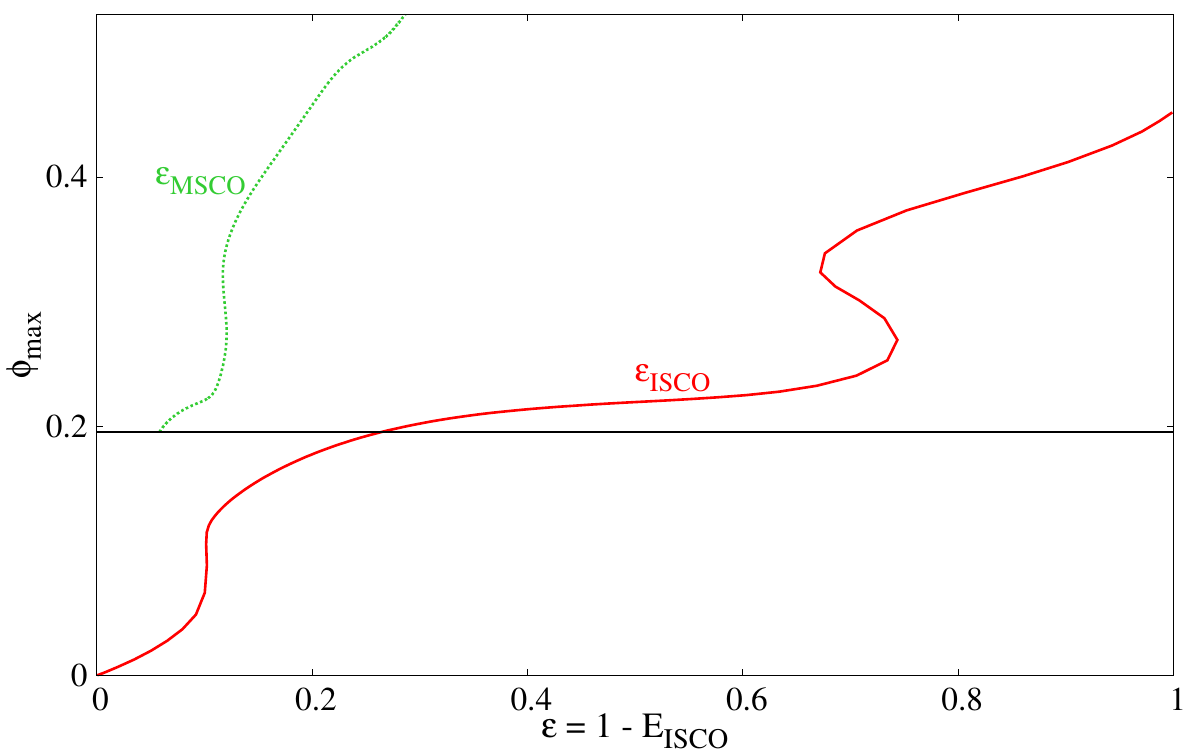}\\
	\includegraphics[scale=0.4]{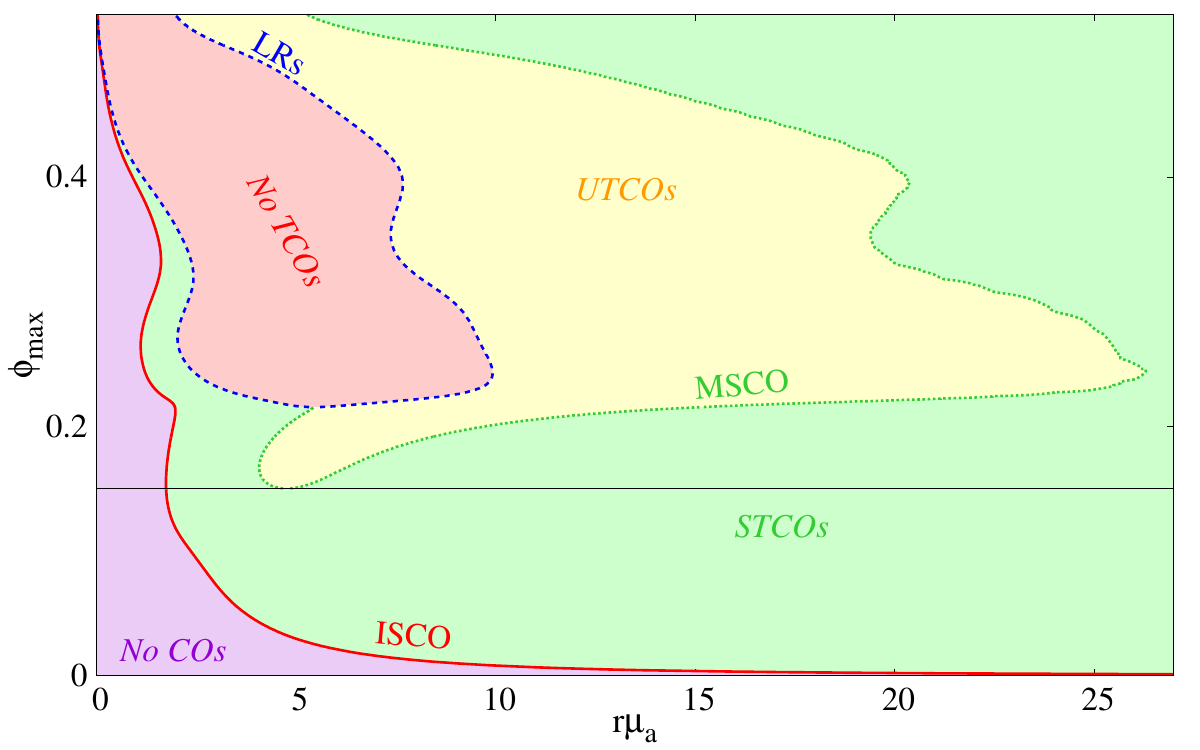}
	\includegraphics[scale=0.4]{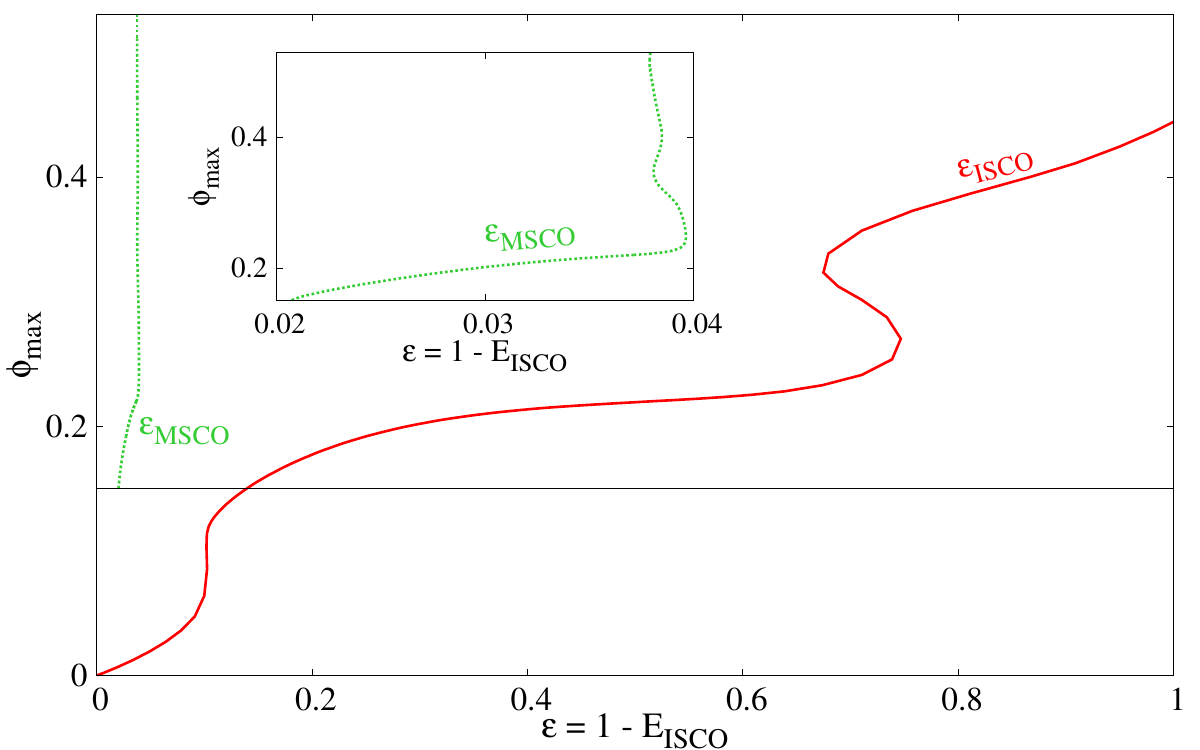}
	\caption{Same as Fig.~\ref{Fig:MiniBS_m_1} but  for axion boson stars with $f_a = 0.03$.}
	\label{Fig:AxionBS_fa_003}
\end{figure}

Further decreasing $f_a$ ($i.e.$ increasing the self-interactions) introduces more convoluted features -  Fig. \ref{Fig:AxionBS_fa_003}. 
This family of axionic stars with $f_a = 0.03$ has the striking feature that $\epsilon_{\rm ISCO}$ is not longer a monotonically increasing function of $\phi_{\rm max}$. In fact there are now solutions with degenerated efficiencies, say with $\phi_{\rm max} = [0.229, 0.376]$. Apart from this novelty, $\epsilon_{\rm ISCO}$ still approaches unity for large $\phi_{\rm max}$.

Concerning $\epsilon_{\rm MSCO}$, in the prograde case, it emerges when $\epsilon_{\rm ISCO}\sim 26\%$ and is $\epsilon_{\rm MSCO}\sim 5.9\%$. Then, it varies non-monotonically: at $\phi_{\rm max} \approx 0.274$, $\epsilon_{\rm MSCO}\sim 12\%$ (local maximum); at $\phi_{\rm max} \approx 0.321$, $\epsilon_{\rm MSCO}\sim 11\%$ (local minimum); for the larger $\phi_{\rm max}$, $\epsilon_{\rm MSCO}\sim 32\%$ (global maximum). 
For retrograde orbits, $\epsilon_{\rm MSCO}$ is more constant. It emerges when $\epsilon_{\rm ISCO}\sim 12\%$, with $\epsilon_{\rm MSCO}\sim 1.2\%$. Then, for $\phi_{\rm  max} \approx 0.247$, it reaches a local maximum, $\epsilon_{\rm MSCO}\sim 4\%$. Going further into the strong gravity regime,  the efficiency is approximately constant.

\subsubsection{Gauged Boson Stars}

Gauged boson stars can be thought as electrically charged mini-boson stars. They are solutions of the (complex-)Einstein-Klein-Gordon-Maxwell theory, 
\begin{equation}
	\mathcal{S} = \int d^4 x \sqrt{-g} \left[ \frac{R}{16\pi} - \frac{1}{4} F_{\mu\nu}F^{\mu\nu} - g^{\mu\nu}D_\mu \Psi^* D_\nu \Psi - \mu^2 \Psi^* \Psi \right] \ ,
\end{equation}
where the electromagnetic tensor $F_{\mu\nu} \equiv \partial_\mu A_\nu - \partial_\nu A_\mu$ is defined through the electromagnetic potential $A_\mu$, and $D_\mu \equiv \partial_\mu + i q_E A_\mu$. There is a minimal coupling between the electromagnetic sector and the scalar field through the (gauge) covariant derivative $D_\mu$, which introduces the gauge coupling constant $q_E$.

The first work on gauged boson stars was develop by Jetzer and van de Bij\cite{jetzer1989charged} where they obtained spherically symmetric solutions -- see also \cite{Pugliese:2013gsa}. The rotating generalisation was constructed  later in a more general context 
in 
\cite{Delgado:2016jxq}, 
(see also 
\cite{Brihaye:2009dxv2,Collodel:2019ohy} for results in a model with a self-interacting scalar field). 

\begin{figure}
	\centering
	\includegraphics[scale=0.4]{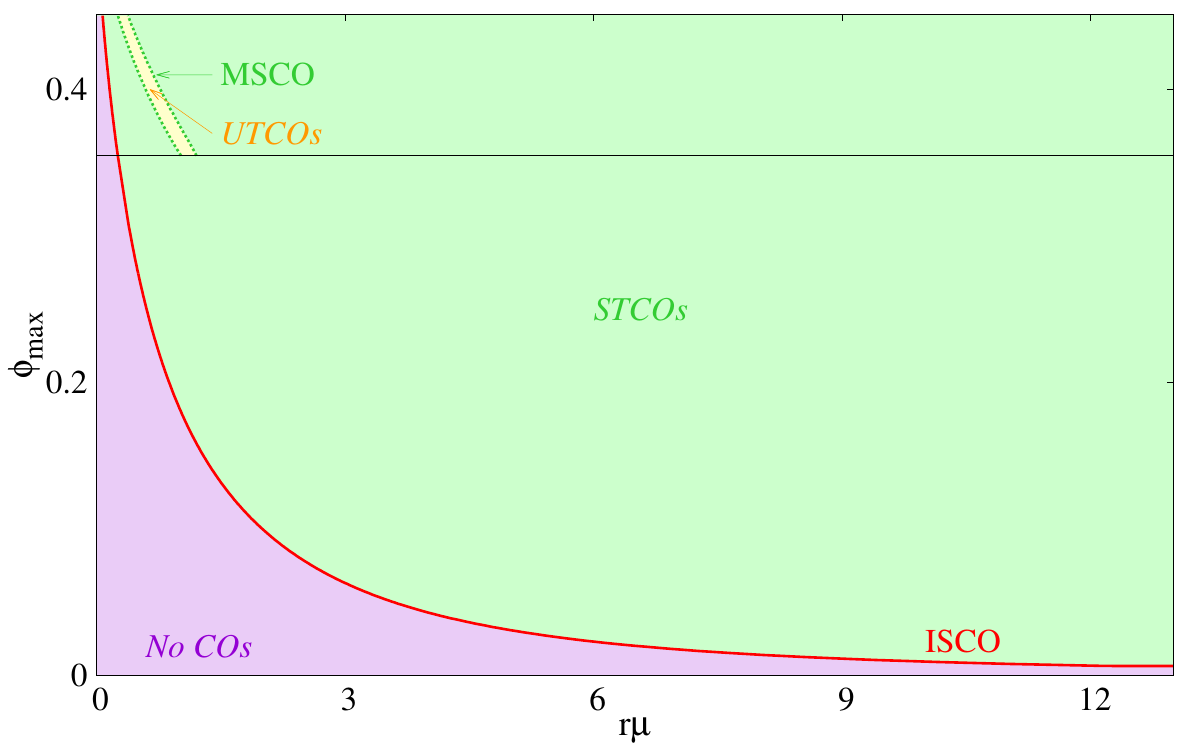}
	\includegraphics[scale=0.4]{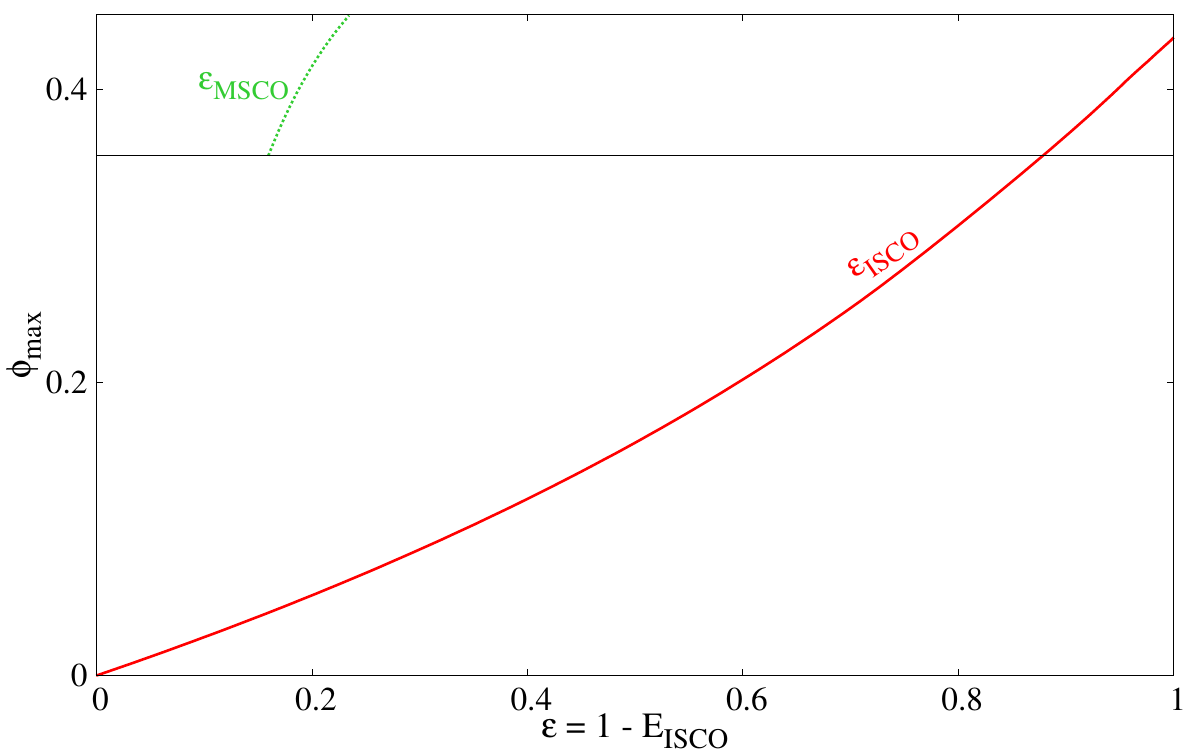}\\
	\includegraphics[scale=0.4]{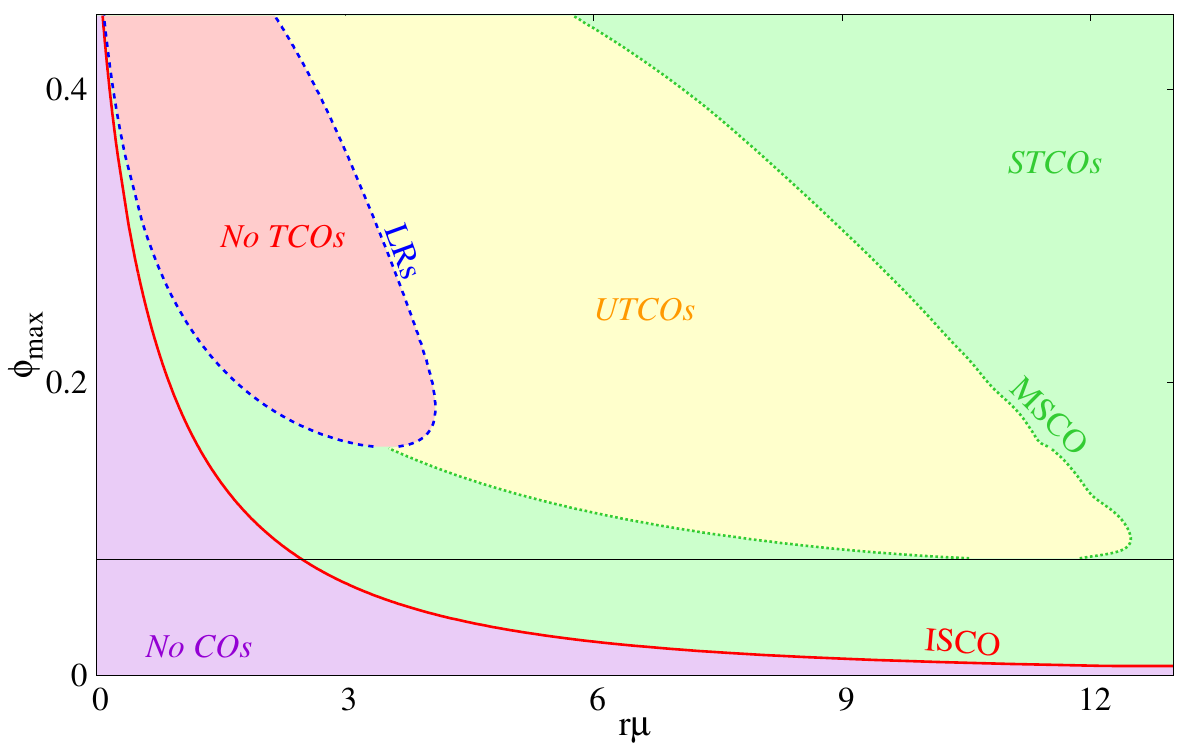}
	\includegraphics[scale=0.4]{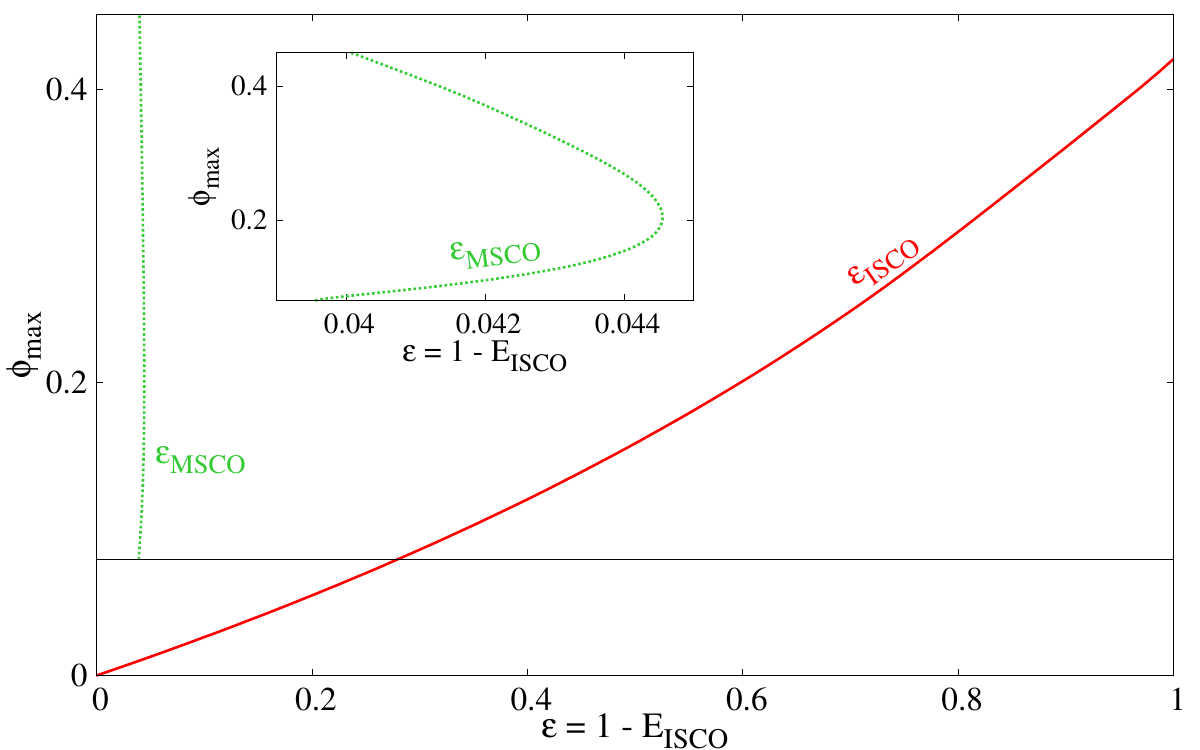}
		\caption{Same as Fig.~\ref{Fig:MiniBS_m_1} but  for gauged boson stars with $q_E = 0.6$.}
	\label{Fig:GaugedBS_qE_06}
\end{figure}

 The latter are found by using the same ansatz as in Eq. (\ref{Eq:AnsatzScalarField}),
together with the $U(1)$ form $A=A_\varphi d\varphi+A_t dt$ \cite{Delgado:2016jxq}.
This implies that, as before, we have to specify the azimuthal harmonic index $m$. Furthermore, we also need to specify the gauge coupling constant $q_E$. In this work we will only consider gauged solution with $m=1$ and $q_E = 0.6$; the latter choice illustrates the generic features we have seen analysing also other values of $q_E$. Such results are shown in Fig. \ref{Fig:GaugedBS_qE_06}. Overall we observe that the description for mini-boson stars with $m=1$ still apply for this case. 
Being more specific, for the prograde case,  $\epsilon_{\rm ISCO}\sim 16\%$ (in contrast to $\sim 25\%$ for mini-boson stars) for the first solution for which $\epsilon_{\rm ISCO}\neq \epsilon_{\rm MSCO}$; then $\epsilon_{\rm MSCO}$ increases gradually until $\sim 24\%$ (in contrast to $\sim 30\%$ for mini-boson stars). For the retrograde case, however, the efficiency difference between gauged and ungauged boson stars is unnoticeable. 

The discussion made here prompts the conclusion that, at least for the gauged boson stars
reported in~\cite{Delgado:2016jxq},  the presence of an electric charge   does not influence significantly
 either the structure of TCOs or the efficiency.

\subsubsection{Proca Stars}
We now consider vector boson stars, known as Proca stars. They are the horizonless and regular everywhere solutions of the (complex)-Einstein-Proca theory, with the following action,
\begin{equation}
	\mathcal{S} = \int d^4 x \sqrt{-g} \left[ \frac{R}{16\pi} - \frac{1}{4} F_{\mu\nu} \bar{F}^{\mu\nu} - \frac{1}{2} \mu_P^2 A_\mu \bar{A}^\mu \right] \ ,
\end{equation}
where $F_{\mu\nu} = \partial_\mu A_\nu - \partial_\nu A_\mu$ is the field strength written in terms of the 4-potential $A_\mu$. The bar over the field strength and 4-potential, $\bar{F}_{\mu\nu}$ and $\bar{A}_\mu$, corresponds to the complex conjugate, while $ \mu_P$ is the vector field mass.  These stars were first studied in \cite{Brito:2015pxa}, where both static and rotating numerical solutions where discussed, together with their physical properties and stability. The spinning solutions therein, however, were excited states. The fundamental spinning solutions were discussed in~\cite{Herdeiro:2019mbz,Santos:2020pmh}.

The Proca potential \textit{ansatz} is,
\begin{equation}
\label{va}
	A = \left(i V dt + \frac{H_1}{r} dr + H_2 d\theta + i H_3 \sin \theta d\varphi \right) e^{i(m \varphi - \omega t)} \ ,
\end{equation} 
where $V, H_1, H_2$ and $H_3$ are functions that only depend on the $(r,\theta)$ coordinates, and $m \in \mathbb{Z}^+$ is the usual azimuthal harmonic index. Given that there is an infinite number of families of Proca stars with different values of $m$, in this work we will only consider the family of $m=1$ Proca stars (and also excited solutions with one node in the radial direction).

\begin{figure}
	\centering
	\includegraphics[scale=0.4]{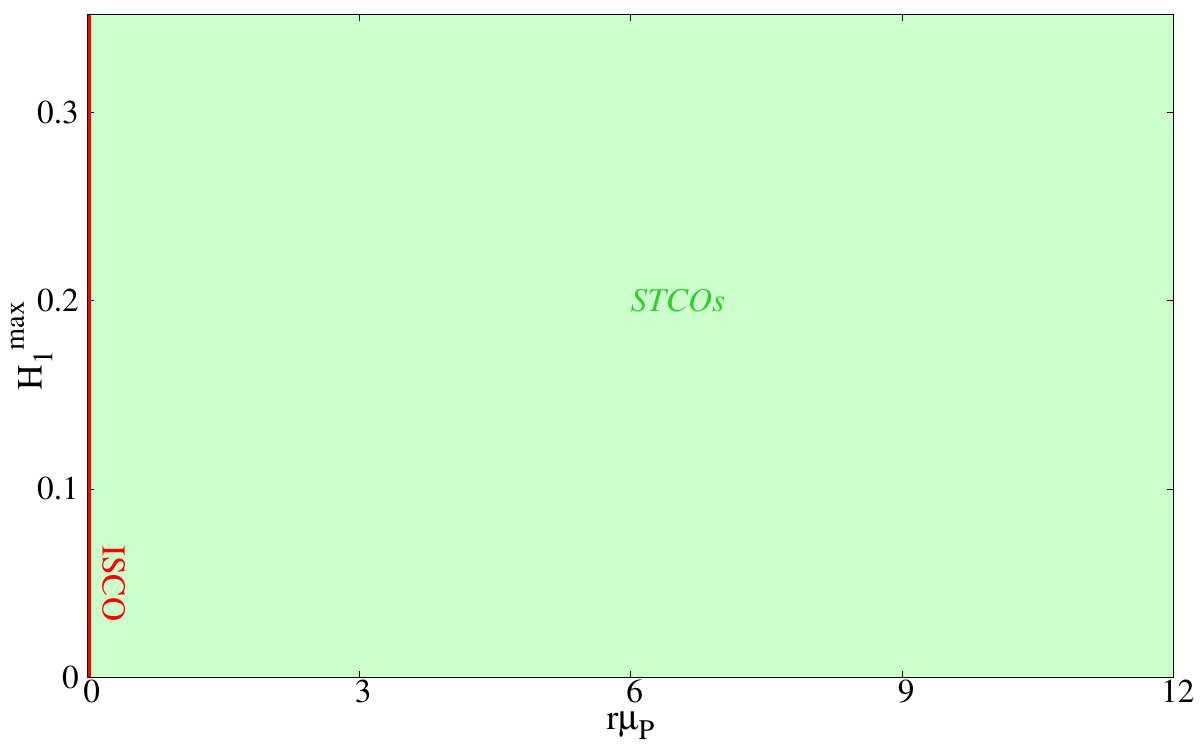}
	\includegraphics[scale=0.4]{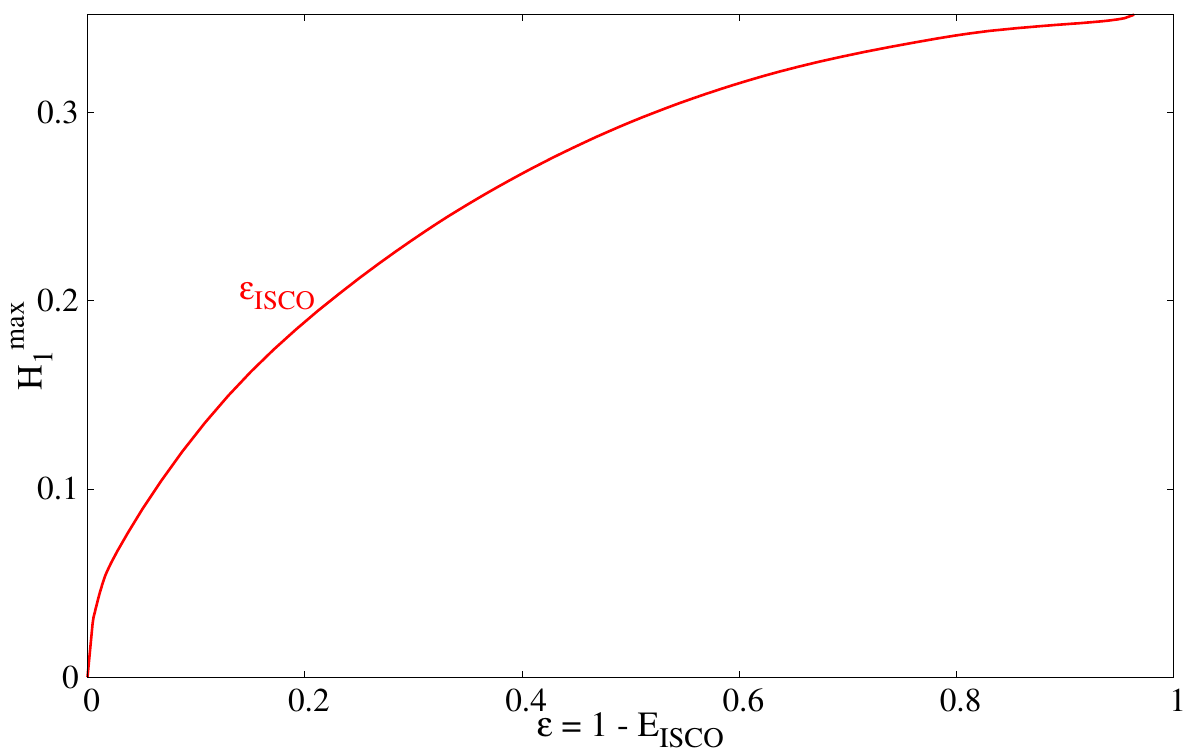}\\
	\includegraphics[scale=0.4]{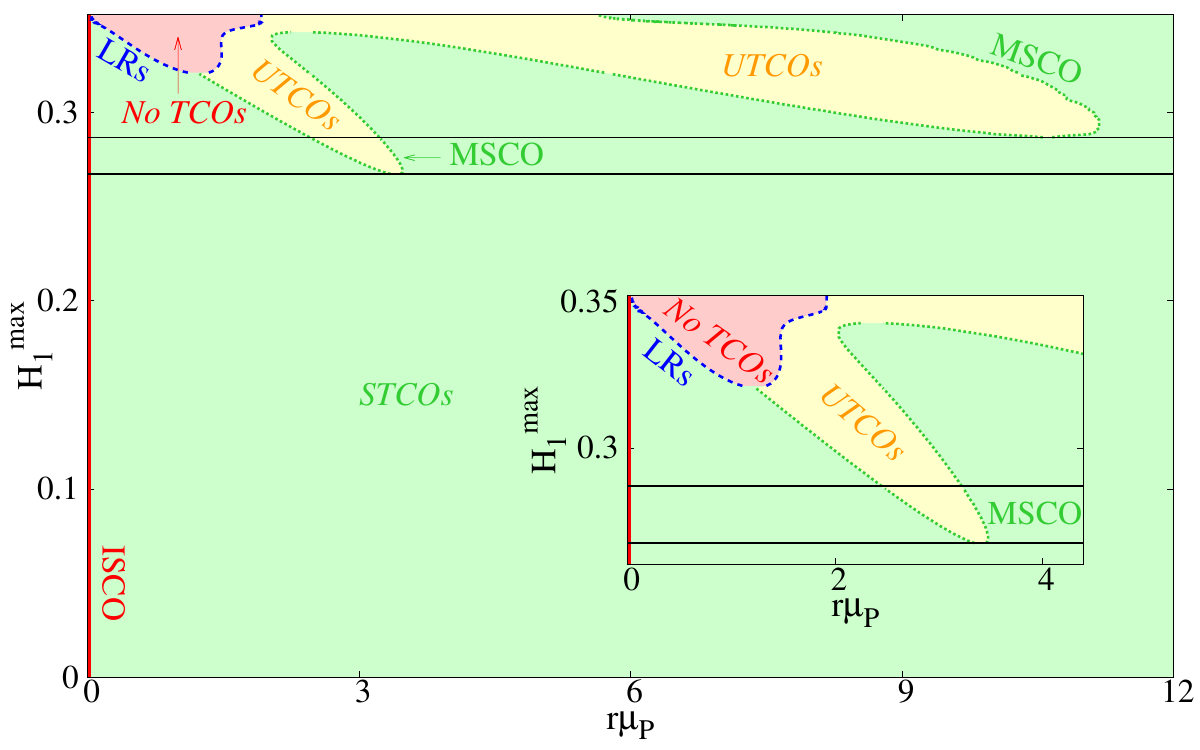}
	\includegraphics[scale=0.4]{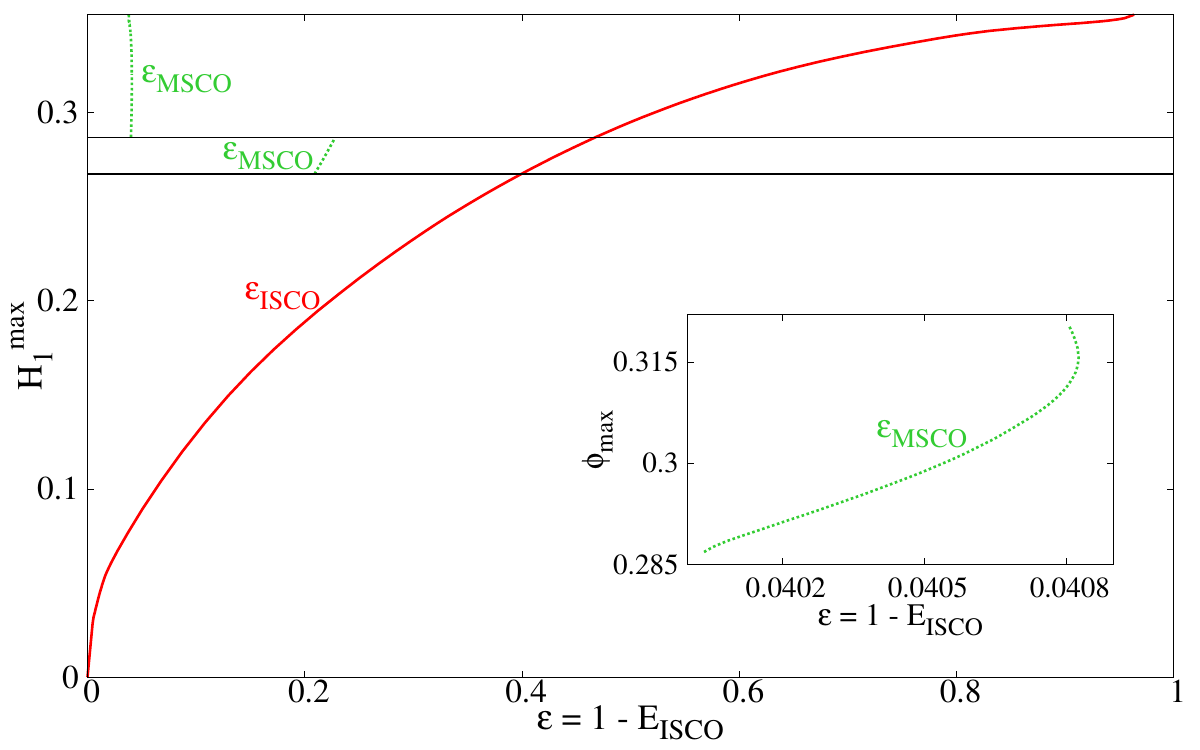}
	\caption{Same as Fig.~\ref{Fig:MiniBS_m_1} but  for Proca stars.}
	\label{Fig:Proca_m_1}
\end{figure}

Since our bosonic field is now a vector, we can no longer use $\phi_{max}$ to label  solutions. This is replaced by the maximal value of the $H_1$ function in the vector ansatz (\ref{va}). 
As for $\phi_{\rm max}$ for the previous stars, the $H_1$ function also increases monotonically  moving from the dilute regime until the strong gravity regime along the domain of existence of Proca stars. Hence, each individual Proca solution has a different value of $H_1^{\rm max}$. Therefore, in Fig. \ref{Fig:Proca_m_1}, the structure of TCOs and LRs is shown in a $H_1^{\rm max}$ \textit{vs.} $r\mu_P$ plot (and similarly for the efficiency $\epsilon$).

Fig. \ref{Fig:Proca_m_1} exhibits clear differences between the vector and scalar stars. A notorious one is the possibility of having  stable TCOs all the way to the center of the star. Thus, ISCO has $r=0$. Then $\epsilon_{\rm ISCO}$ amounts to known the energy of the particle sitting at $r=0$. We see that such $\epsilon_{\rm ISCO}$ goes from zero until close to unity as more compact stars are considered, similarly to the scalar stars.

Concerning $\epsilon_{\rm MSCO}$, the sense of rotation plays a role. For prograde orbits, there is no region of  unstable TCOs. Thus $\epsilon_{\rm ISCO}=\epsilon_{\rm MSCO}$. For retrograde orbits, however,  unstable TCOs can appear. In fact, there can even be several disconnected regions of such orbits. Thus $\epsilon_{\rm MSCO}$ presents more than one discontinuity. The first discontinuity appears when $\epsilon_{\rm ISCO}\sim 40\%$, and $\epsilon_{\rm MSCO}\sim 21\%$. Then, $\epsilon_{\rm MSCO}$ increases slightly up to $\epsilon_{\rm MSCO}\sim 23\%$ where the second discontinuity appears, dropping further to $\epsilon_{\rm MSCO}\sim 4\%$. For the remaining solutions on the strong gravity regime, it decreases slowly down to $\epsilon_{\rm MSCO}\sim 3.8\%$.

\subsection{BHs}

Now we consider (non-Kerr) BH examples. Our illustrations, again, are numerical. The line element considered in this work, which is common for the two families of BHs discussed below, is,
\begin{equation}
	ds^2 = -e^{2F_0} N + e^{2F_1} \left( \frac{dr^2}{N}  + r^2 d\theta^2 \right) + e^{2F_2} r^2 \sin^2 \theta \left( d\varphi - W dt \right)^2~, \hspace{10pt} N \equiv 1 - \frac{r_H}{r} \ ,
\end{equation}
where $r_H$ is the radial coordinate of the event horizon and, as for the stars case, $F_0, F_1, F_2$ and $W$ are ansatz functions that depends only on $(r,\theta)$.

For the case of BHs, we choose to show the efficiency as a function of the dimensionless spin, $j = J/M^2$ in the two plots for each family of solutions. The left (right) plot corresponds to prograde orbits (retrograde orbits).

\subsubsection{BHs with Synchronised Axionic Hair}

To illustrate the structure of TCOs and LRs for a non-Kerr family of BHs that can exhibit large phenomenological deviations  from Kerr we consider BHs with synchronised hair. We will consider the axionic model~\cite{Delgado:2020hwr}, which contains in a particular limit the free scalar field model~\cite{Herdeiro:2014goa,Herdeiro:2015gia}. Some results  for the latter, concerning the efficiency, were recently  presented in~\cite{Collodel:2021gxu}.

BHs with synchronised axionic hair are stationary, axisymmetric, regular everywhere on and outside the event horizon, asymptotically flat solutions of the (complex-)Einstein-Klein-Gordon theory -- \textit{cf.} Eq. (\ref{Eq:ActionAxion}). 
They can be consider as the natural BH generalisation of the axion boson stars studied previously, thus the self-interaction potential $V(\phi)$ follows the QCD axion potential, similar as for stars case -- \textit{cf.} Eq. (\ref{Eq:AxionPotential}). 
In this work we will only consider the BHs generalisation of the axion boson stars with $f_a = 0.05$.

BHs with synchronised axionic hair are composed of a BH horizon surrounded by an axionic scalar field whose angular frequency is synchronised with the angular rotation of the horizon of the BHs. Such synchronisation can be written as $\omega = m \Omega_H$, where $\Omega_H$ is the angular velocity of the horizon. If this synchronisation is not met, the scalar field cannot be in equilibrium with the BH.

\begin{figure}[h!]
	\centering
	\includegraphics[scale=0.4]{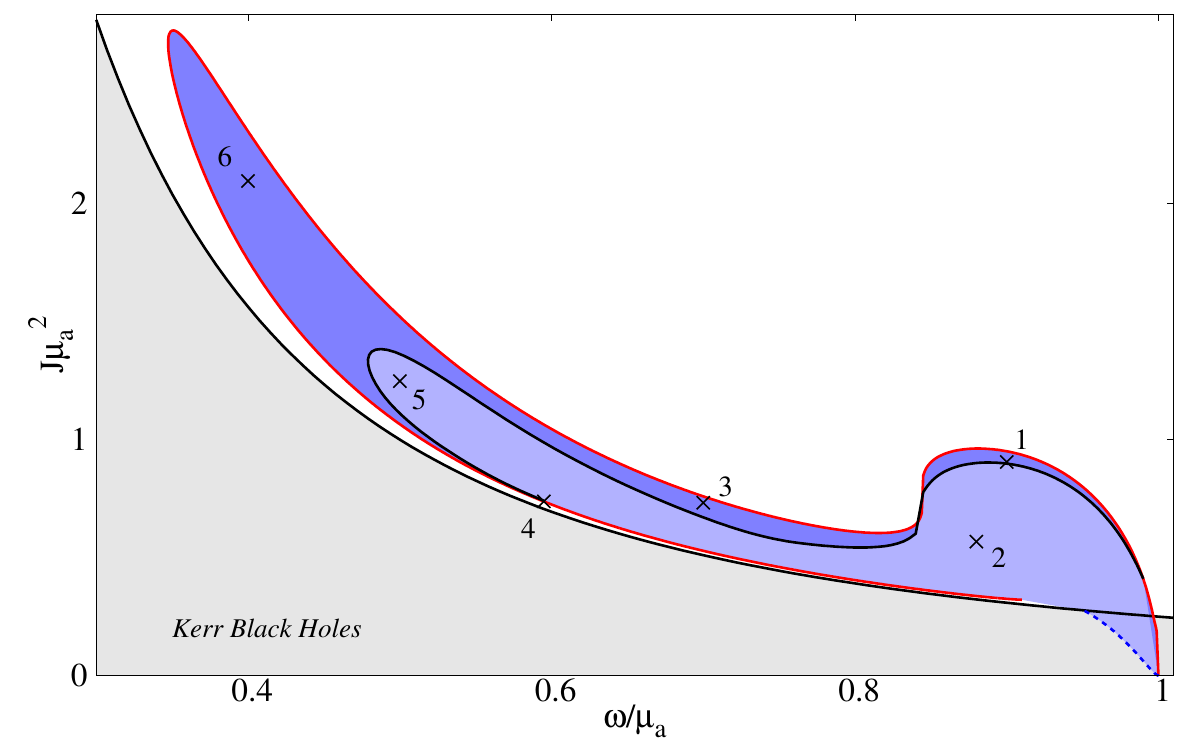}
	\includegraphics[scale=0.4]{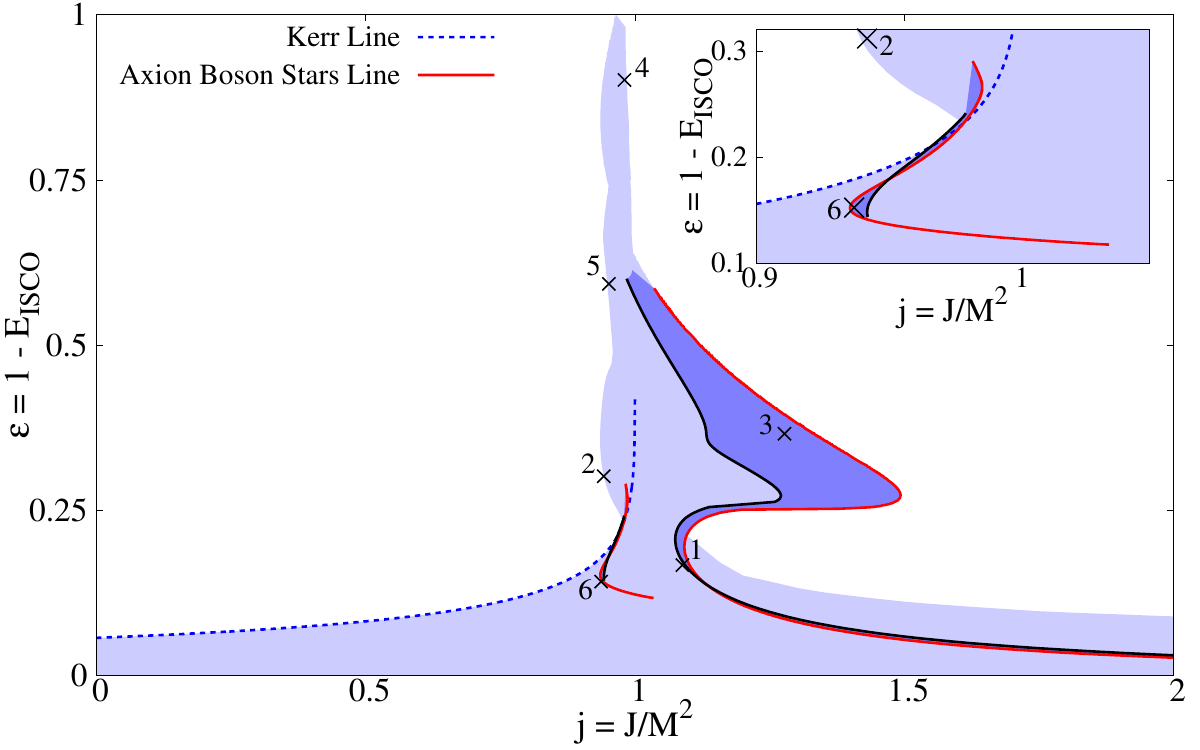}\\
	\includegraphics[scale=0.4]{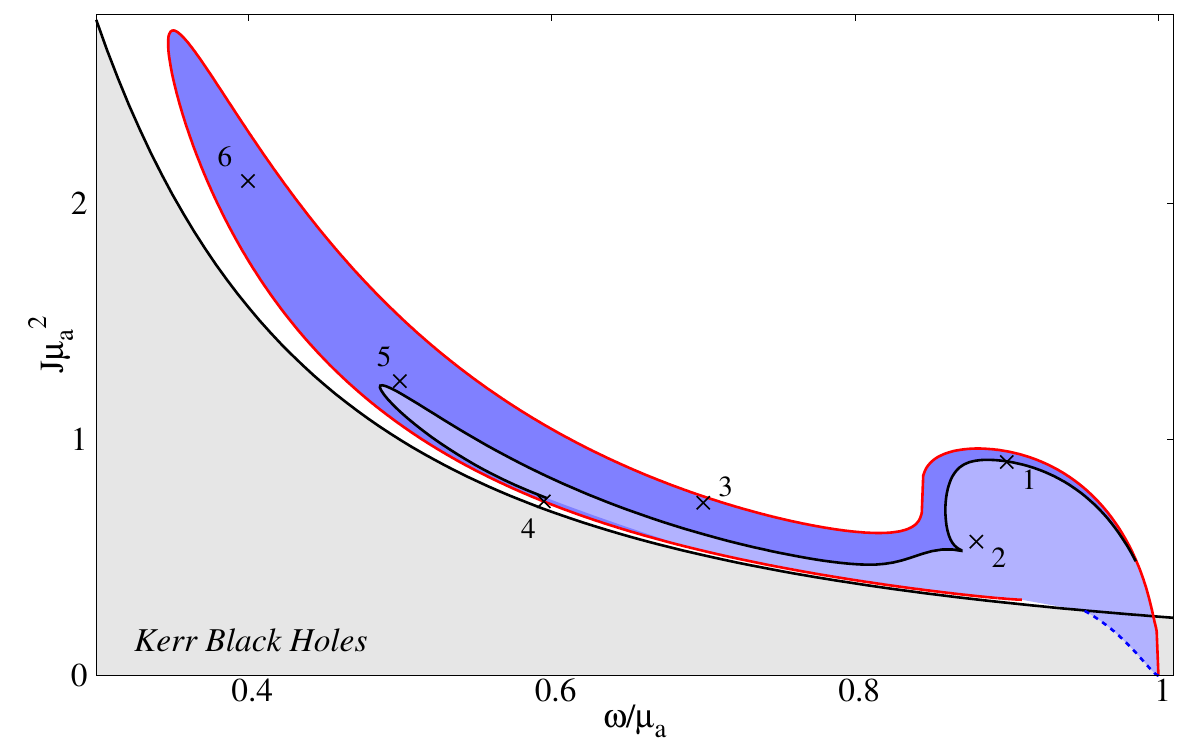}
	\includegraphics[scale=0.4]{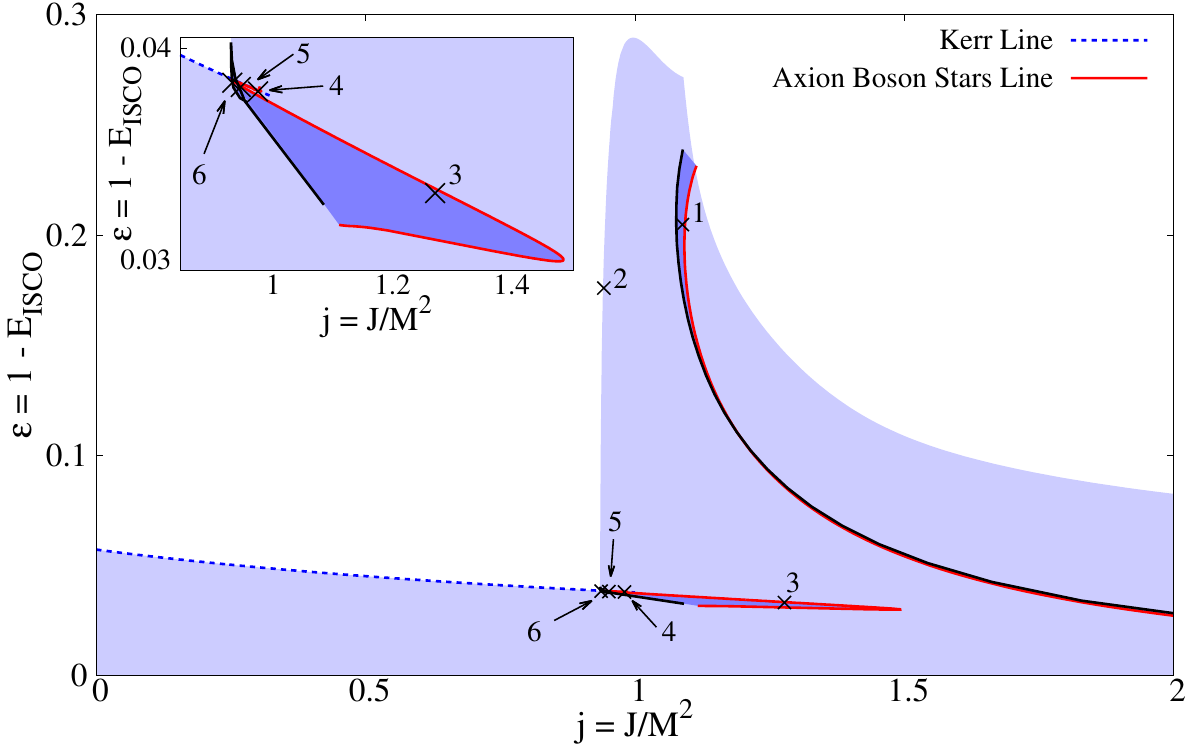}
	\caption{Domain of existence in an angular momentum $J\mu_a^2$ \textit{vs.} angular frequency of the scalar field $\omega/\mu_a$ plane (\textit{left column}) and efficiency as a function of the dimensionless spin $j = J/M^2$ (\textit{right column}) of BHs with synchronised axionic hair with $f_a=0.05$. prograde (retrograde) orbits are presented in the \textit{top row} (\textit{bottow row}). The light blue regions correspond to solutions in which the efficiency is $\epsilon_{\rm ISCO}$; the dark blue regions correspond to solutions for which the efficiency is $\epsilon_{\rm MSCO}$. The inset in both efficiency plots sheds more light onto the regions that are more challenging to analyse on the main plots. Six solutions are highlighted.}
	\label{Fig:KBHsASH}
\end{figure}

This new family of BHs was first obtained in \cite{Delgado:2020hwr}, where the authors also studied some physical properties of the solutions, as well as some phenomenological proprieties, including the structure of TCOs, which can be quite different from the one for Kerr. For solutions with a small amount of hair, the structure is similar to the Kerr one. There is only one  unstable LR. Between the event horizon and the LR, there are no TCOs. Between the LR and the ISCO, there are  unstable TCOs; and above the ISCO, there are  stable TCOs. Thus, for these solutions with a small amount of hair, $\epsilon_{\rm ISCO}=\epsilon_{\rm MSCO}$. The results for solutions in this class (Kerr-like) are represented in  light blue for  both plots in Fig. \ref{Fig:KBHsASH}. 
On the other hand, very ``hairy" solutions  present a more convoluted structure of TCOs. There are new disconnected regions of  unstable TCOs and forbidden for TCOs, where the latter appear when the scalar field is compact enough to develop extra LRs. Hence, we can have $\epsilon_{\rm ISCO}\neq \epsilon_{\rm MSCO}$. The results for solutions in this class (non-Kerr like) are represented in dark blue color for both plots in Fig. \ref{Fig:KBHsASH}.

In Fig. \ref{Fig:KBHsASH} we show $\epsilon_{\rm ISCO}$ for BHs with synchronised axionic hair for prograde (\textit{left}) and retrograde (\textit{right}) orbits. Both panels also include an inset plot showing the domain of existence of these BHs in an angular momentum $J\mu_a^2$ \textit{vs.} $\omega/\mu_a$ diagram. Both in the main panels and insets, there are two additional lines. The first one corresponds to no horizon limit: the set of axion boson stars with  $f_a = 0.05$. This (red solid) line is known as the \textit{Axion Boson Stars line}. The second (blue dashed) line corresponds to the no hair limit or Kerr limit of the hairy BHs - the \textit{Kerr line}. We have highlighted six particular solutions, numbered 1 to 6, to allow an easier mapping between the domain of existence and the efficiency plot.

 Fig. \ref{Fig:KBHsASH} (left panel) shows there are prograde efficiencies $\epsilon_{\rm ISCO}$ arbitrarily close to the unity, exceeding greatly the maximal efficiency of $\sim 42\%$ (of the Kerr limit). The solutions with the largest efficiencies correspond to solutions in the strong gravity regime, where the ISCO occurs for smaller radii, leading to larger $\epsilon_{\rm ISCO}$. For non-Kerr like solutions -- dark blue region -- $\epsilon_{\rm MSCO}$ can be as high as $\sim 60\%$, again, larger than the maximal efficiency for Kerr BHs.  For very non-Kerr like solutions,  $\epsilon_{\rm MSCO}$ drops to around $\sim 20\%$ because a new region of  unstable TCOs develops, pushing MSCO outwards.

Fig. \ref{Fig:KBHsASH} (right panel) addresses the retrograde case.  Efficiencies are rather smaller than in the prograde one, since both ISCO and MSCO
occur at larger radii. Additionally, a new disconnected regions of (no or  unstable) TCOs develop for solutions with far less hair than for the previous case, pushing MSCO outwards. Nevertheless, retrograde efficiencies of  $\epsilon_{\rm ISCO}\sim 30\%$ are possible, far larger than those for the retrograde case in Kerr BHs.

\subsubsection{Einstein-scalar-Gauss-Bonnet BHs}

The goal of our final example is to stress that, in fact, many models of non-Kerr BHs have small phenomenological differences with respect to the  Kerr model. In particular this applies to the efficiencies we have been discussing. We will discuss Einstein-scalar-Gauss-Bonnet BHs, which are asymptotically flat, regular everywhere outside and at the event horizon, axisymmetric and stationary solutions of the Horndeski shift-symmetric theory. This is a scalar-tensor theory, within the generic class of Einstein-scalar-Gauss-Bonnet models given by
\begin{equation}
	\mathcal{S} = \int d^4 x \sqrt{-g} \left[ R - \frac{1}{2} \partial_\mu \phi \partial^\mu \phi + \alpha f(\phi) R_{\text{GB}}^2 \right] \ ,
\end{equation}
where $\alpha$ is a dimensionful coupling constant, $f(\phi)$ is a dimensionless coupling function, and $R_{\text{GB}}^2 \equiv R_{\alpha\beta\mu\nu}R^{\alpha\beta\mu\nu} - 4 R_{\mu\nu}R^{\mu\nu} + R^2$ is the well known Gauss-Bonnet quadratic curvature invariant. This class of models is, itself, a subclass of all possible scalar-tensor theories with second order equation of motion -- Horndeski theory \cite{horndeski1974second}. To further specify which scalar-tensor theory we will address, we impose that the coupling function is a linear function of $\phi$,
\begin{equation}
	f(\phi) = \phi \ .
\end{equation}
This choice implies that this theory is shift symmetric, \textit{i.e.}, it is invariant under transformations of the type,
\begin{equation}
	\phi \rightarrow \phi + \phi_0 \ ,
\end{equation}
where $\phi_0$ is an arbitrary constant. This follows from the fact that, in four spacetime dimensions, the Gauss-Bonnet term alone is a total divergence. 

BHs solution within this theory were first obtained by Sotiriou and Zhou \cite{Sotiriou:2013qea,Sotiriou:2014pfa}. In their work, they first showed that the existing no-scalar-hair theorem for this theory can be circumvented since one of the assumptions of the theorem (finiteness of a certain current) can be violated. Then, they obtained analytically, static perturbative solutions (small values of $\alpha$), as well as  numerical static solutions (large values of $\alpha$). A similar work for the spinning generalisation of these solutions was reported in \cite{Delgado:2020rev}.

The structure of TCOs (not shown here) is always Kerr-like. Thus, similarly to the solutions with small amounts of hair in the previous family of hairy BHs, $\epsilon_{\rm ISCO}=\epsilon_{\rm MSCO}$.

\begin{figure}[h!]
	\centering
	\includegraphics[scale=0.43]{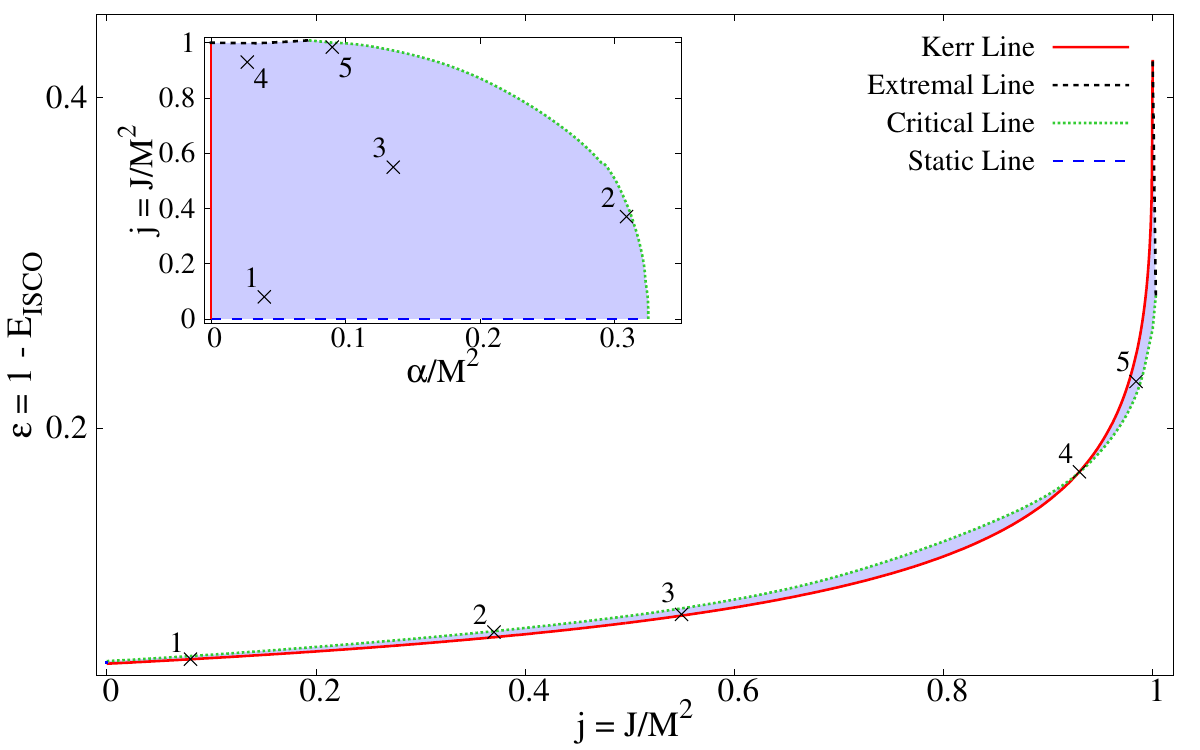}
	\includegraphics[scale=0.43]{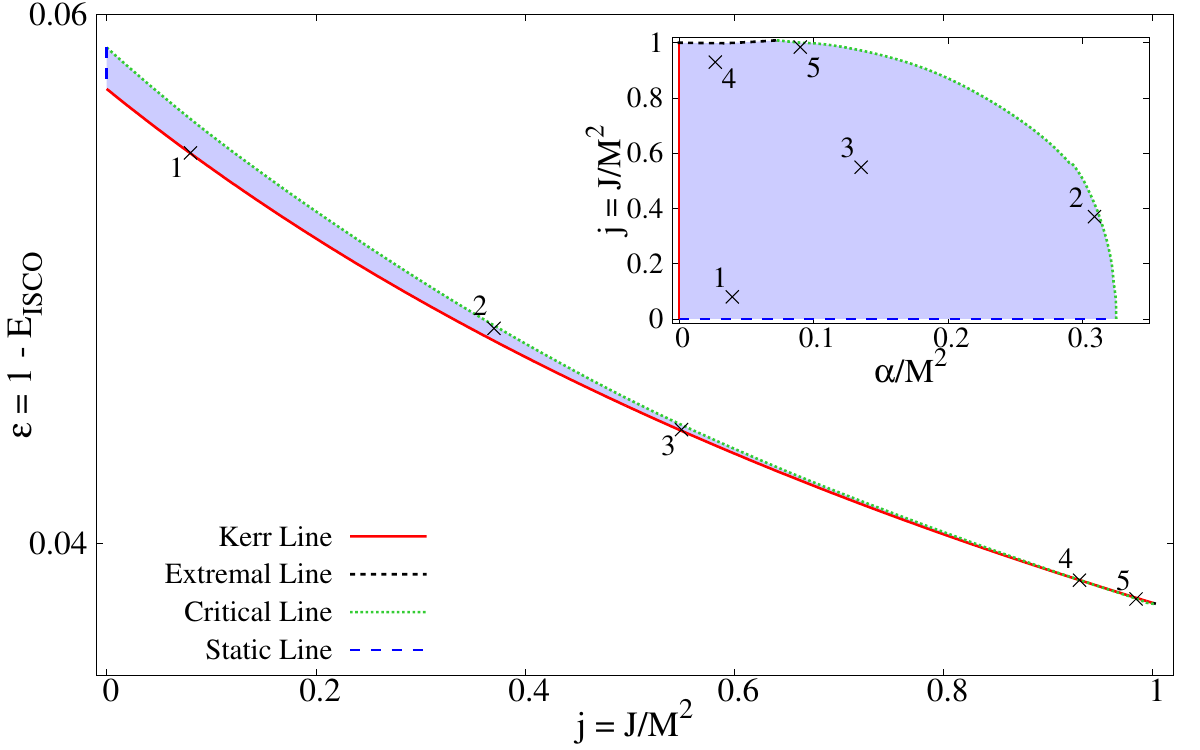}
	\caption{Efficiency as a function of the dimensionless spin $j = J/M^2$ for Einstein-scalar-Gauss-Bonnet BHs. The \textit{left plot} (\textit{right plot}) exhibits prograde (retrograde) orbits. The insets exhibit the domain of existence of Einstein-scalar-Gauss-Bonnet BHs in a dimensionless spin $j = J/M^2$ \textit{vs.} $\alpha/M^2$ plane. Five solutions are highlighted.} 
	\label{Fig:Horndeski}
\end{figure}

In Fig.~\ref{Fig:Horndeski} we show the efficiency for Einstein-scalar-Gauss-Bonnet BHs for prograde (\textit{left}) and retrograde (\textit{right}) orbits. We also include insets  showing the domain of existence of these BHs in a dimensionless spin, $j = J/M^2$ \textit{vs} $\alpha/M^2$ plot. Four additional lines are exhibited. 
The first (green dotted) line is known as the \textit{critical line}  and corresponds to the  limit  beyond which the (repulsive) Gauss-Bonnet term  prevents the existence of a horizon. 
The second (black dashed) line corresponds to the set of extremal hairy solutions with a vanishing Hawking temperature - the \textit{extremal line}. 
The third (blue dashed) line corresponds to non-rotating BHs -  the \textit{static line}. 
Finally, a fourth (solid red) line corresponds to Kerr BHs, in which $\alpha/M^2 = 0$ -  the \textit{Kerr line}. Five particular solutions, numbered  1 to 5, are also highlighted, to map their location in the domain of existence and in the efficiency plot. 

Fig. \ref{Fig:Horndeski} (left panel) shows that the efficiency of Einstein-scalar-Gauss-Bonnet BHs is very similar to the efficiency of Kerr BHs, for  the same dimensionless spin, $j$. The largest difference  is (only) around $\sim 4\%$. For small $j$, the Einstein-scalar-Gauss-Bonnet BHs have a larger efficiency; but, for sufficiently large $j$, the reverse happens. This sort of transition was already discussed in \cite{Delgado:2020rev} (albeit not for the efficiency).
%
The right panel in Fig. \ref{Fig:Horndeski} exhibits a similar picture  for retrograde orbits.  The largest difference is now (only) around $\sim 3\%$ and occurs in the static limit, $j \rightarrow 0$. Increasing the spin, this difference monotonically decreases. For large spins, there is almost no difference between hairy and Kerr BHs in terms of efficiency. This result is consistent with the discussion in~\cite{Delgado:2020rev}.

\section{Discussion and final remarks}

In this work, we have shown that for stationary, axisymmetric and asymptotically flat compact objects with a $\mathbb{Z}_2$ symmetry, the existence of equatorial LRs leads to a specific structure for equatorial TCOs, independently of the direction of rotation. Such structure is entirely determined by the  stability of the LR: for an  unstable LR, the region  radially  immediately above (below) the LR has  unstable TCOs (no TCOs) -- \textit{cf.} Fig. \ref{Fig:CompIllustStableLR} (top panel); for a stable LR, the region radially immediately  above (below) the LR has no TCOs (has  stable TCOs) -- \textit{cf.} Fig. \ref{Fig:CompIllustStableLR} (bottom  panel).

As a corollary of this result, for a horizonless object that possesses one  unstable LR and another  stable LR at a smaller radius than the first, for either sense of rotation, the region between the LRs has no TCOs -- \textit{cf.} Fig.~\ref{Fig:CompIllustBHs} (top panel). Radially immediately above (below) the unstable (stable) LR, there are  unstable (stable) TCOs.  This implies that it is possible to have  stable TCOs closer to the object itself than the LR; thus, a potential ISCO may occur at a smaller radius than the LR.
However, one needs to clarify if the motion on such region is perturbatively stable in a 
direction perpendicular to the equatorial plane.

As another corollary, for asymptotically flat equilibrium BHs, which generically have an unstable  LR for either rotation sense \cite{Cunha:2020azh}, the region between the event horizon and the unstable LR contains no TCOs -- \textit{cf.} Fig.~\ref{Fig:CompIllustBHs} (bottom panel). Since the LR is unstable, the region radially immediately above has  unstable TCOs; thus, for a BH, the ISCO will always occur at a larger radius than this unstable LR. 

In the second part of this work, we have studied the efficiency associated to the process of converting gravitational energy into radiation by a material particle falling under an adiabatic sequence of TCOs, for several stars and BHs, namely, three different families of bosonic scalar stars (mini, gauged and axion boson stars), one family of bosonic vector (Proca) stars  and two different families of hairy BHs (BHs with synchronised axionic hair and Einstein-scalar-Gauss-Bonnet BHs). 

Regarding the several families of bosonic scalar stars, we found that the structure of TCOs is quite similar between them.
Moreover, their  structure is also similar to that found for some naked singularities -- see Refs.
 \cite{Pugliese:2011py,Stuchlik:2002tj,Stuchlik:2014yaa,Vieira:2013jga}. 
The efficiency $\epsilon_{\rm ISCO}$ computed at the ISCO, can grow arbitrarily close to unity, both for co- and retrograde orbits.
Also, the efficiency $\epsilon_{\rm MSCO}$, at the MSCO, has the largest values for both stars without self-interactions (mini and gauged boson stars) and prograde orbits.

The family of bosonic vector stars presents a structure of TCOs quite different from their bosonic scalar cousins. For prograde orbits,  stable TCOs can exist all the way until $r=0$. Thus $\epsilon_{\rm ISCO}$ is computed at the origin and it increases monotonically towards values close to $100\%$ for stars in the strong gravity regime. For retrograde orbits, more compact stars develop regions of  unstable and no TCOs; thus, the efficiency $\epsilon_{\rm MSCO}$ drops to small values, around $\sim 4\%$.

For BHs with synchronised axionic hair, we found that new disconnected regions of  unstable and no TCOs (beside the ones that exist already for Kerr BHs) develop. Thus, the efficiency $\epsilon_{\rm MSCO}$ can drop; nevertheless, it is possible to have solutions in which this efficiency for prograde orbits is much larger than the one for Kerr BHs and even close to the unity. In the case of retrograde orbits, the efficiency  can not be as high, but  can, nonetheless, be higher than that for the (retrograde) Kerr case.

Finally, concerning the family of Einstein-scalar-Gauss-Bonnet BHs we found that the higher-order correction to Einstein's gravity which arise from the linear coupling between the Gauss-Bonnet term and the scalar field has no strong influence on the efficiency. For prograde orbits, the efficiency is only slightly larger (smaller) than that of Kerr BHs for the same $j$, when $j
$ is small (large). For retrograde orbits, the efficiency of Einstein-scalar-Gauss-Bonnet BHs is larger than their Kerr counterpart, but the difference decreases almost to zero as $j$  increases.

The study in this paper could also relate to gravitational waves. Since the efficiency was initially associated with the study of quasars in the centre of supermassive BHs, where the infalling matter could create an extremely luminous active galactic nucleus, a natural follow up work would be the study of Extreme Mass Ratio Inspirals (EMRIs) for the several families of solutions studied in this work. The infalling particles will have a gravitational wave signal that will chirp up to a cutoff frequency, which is determined by the angular velocity of timelike particles at the ISCO. The results found in such a study could, in principle, be used by the LISA collaboration in the future to help test the Kerr hypothesis. In fact, as a teaser,  a recent study showed that Kerr BHs with scalar hair~\cite{Herdeiro:2014goa} can lead to very different EMRIs than the Kerr geometry~\cite{Collodel:2021jwi}.

\section*{Acknowledgements}

We would like to thank the anonymous referee for the comments and references provided. We would like to also thank P. V. Cunha for comments on a draft of this paper. J. D. is supported by the FCT grant SFRH/BD/130784/2017. This work is supported by the Center for Research and Development in Mathematics and Applications (CIDMA) through the Portuguese Foundation for Science and Technology (FCT - Funda\c c\~ao para a Ci\^encia e a Tecnologia), references UIDB/04106/2020 and UIDP/04106/2020 and by national funds (OE), through FCT, I.P., in the scope of the framework contract foreseen in the numbers 4, 5 and 6 of the article 23, of the Decree-Law 57/2016, of August 29, changed by Law 57/2017, of July 19. We acknowledge support from the projects PTDC/FIS-OUT/28407/2017,  CERN/FIS-PAR/0027/2019 and PTDC/FIS-AST/3041/2020. This work has further been supported by the European Union's Horizon 2020 research and innovation (RISE) programme H2020-MSCA-RISE-2017 Grant No.~FunFiCO-777740. The authors would like to acknowledge networking support by the COST Action CA16104.

\appendix

\section{Circular Motion of Spacelike Geodesics}
\label{App:AppendixA}

Consider the same assumptions and symmetries discussed at the beginning of Section \ref{Sec:Sec2}. In such spacetime, the effective Lagrangian of a spacelike test particle can be written as,
\begin{equation}
	2\mathcal{L} = g_{\mu\nu} \bar{x}^\mu \bar{x}^\nu = 1~,
\end{equation}
where the bar denotes the derivative with respect to arc length. 
Assuming that the motion occurs on the equatorial plane, $\theta = \pi/2$, we can write the Lagrangian as,
\begin{equation}
	2\mathcal{L} = g_{tt}(r,\theta=\pi/2) \bar{t}^2 + 2 g_{t\varphi}(r,\theta=\pi/2) \bar{t} \bar{\varphi} + g_{rr}(r,\theta=\pi/2) \bar{r}^2 + g_{\varphi\varphi}(r,\theta=\pi/2) \bar{\varphi}^2 = 1~.
\end{equation}
Hereafter we will drop the radial dependence of the metric functions to simplify the notation. Due to stationarity and axial-symmetry, we can introduce the energy and angular momentum of the spacelike particle,
\begin{equation}
	-E \equiv g_{t\mu} \bar{x}^\mu = g_{tt} \bar{t} + g_{t\varphi} \bar{\varphi}~, \hspace{10pt} L \equiv g_{\varphi\mu} \bar{x}^\mu = g_{t\varphi} \bar{t} + g_{\varphi\varphi} \bar{\varphi} \ .
\end{equation}
Rewriting the Lagrangian with these new quantities,
\begin{equation}
	2\mathcal{L} = - \frac{A(r,E,L)}{B(r)} + g_{rr} \bar{r}^2 = 1  \ ,
\end{equation}
where, similar as before, $A(r,E,L) = g_{\varphi\varphi} E^2 + 2 g_{t\varphi} E L + g_{tt} L^2$ and $B(r) = g_{t\varphi}^2 - g_{tt} g_{\varphi\varphi}$.
We can now introduce the potential $V_1(r)$ as,
\begin{equation}
	V_1(r) \equiv g_{rr} \bar{r}^2 = 1 + \frac{A(r,E,L)}{B(r)} \ .
\end{equation}
To have a particle following a circular orbit at $r = r^\text{cir}$, both the potential and its radial derivative must be null, hence,
\begin{equation}
	V_1(r^\text{cir}) = 0 \hspace{10pt} \Leftrightarrow \hspace{10pt} A(r^\text{cir},E,L) = - B(r^\text{cir}) \ ,
\end{equation}
and
\begin{equation}
	V_1'(r^\text{cir}) = 0 \hspace{10pt} \Leftrightarrow \hspace{10pt} A'(r^\text{cir},E,L) = - B'(r^\text{cir}) \ .
\end{equation}
Along such circular orbit, the angular velocity of the particle (measured by an observer at infinity) is,
\begin{equation}
	\Omega = \frac{d\varphi}{dt} = \frac{\bar{\varphi}}{\bar{t}} = - \frac{E g_{t\varphi} + L g_{tt}}{E g_{\varphi\varphi} + L  g_{t\varphi}} \ .
\end{equation}

Solving the equation $V_1(r^\text{cir}) = 0$ together with the equation for the angular velocity, we can write the energy and angular momentum of the spacelike particle,
\begin{equation}
	E_\pm =  - \left. \frac{g_{tt} + g_{t\varphi} \Omega_\pm}{\sqrt{-\beta_\pm}} \right|_{r^\text{cir}}~, \hspace{10pt} L_\pm = \left. \frac{g_{t\varphi} + g_{\varphi\varphi} \Omega_\pm}{\sqrt{-\beta_\pm}} \right|_{r^\text{cir}} \ ,
\end{equation}
where $\beta_\pm \equiv (-g_{tt} - 2 g_{t\varphi}\Omega_\pm - g_{\varphi\varphi} \Omega_\pm^2)|_{r^\text{cir}} = - A(r^\text{cir},\Omega,\Omega)$ is the same function defined for the timelike particle case, Eq. \ref{betapm}.

Solving the second equation, $V_1'(r^\text{cir}) = 0$, together with the previous results, we can compute the angular velocity of the spacelike particle,
\begin{equation}\label{Eq:AngVelSpaceLike}
	\Omega_\pm = \left[ \frac{-g_{t\varphi}' \pm \sqrt{C(r)} }{g_{\varphi\varphi}'} \right]_{r^\text{cir}}~.
\end{equation}
This is the same expression for the angular velocity as we saw for timelike particles, Eq. \ref{Eq:AngVelTimeLike}.

From these results we can conclude that when circular orbits are possible, \textit{i.e.} $C(r) \geqslant 0$, the only difference between the circular motion of timelike and spacelike particles resides on the energy and angular momentum, or more precisely, on their dependency with the $\beta_\pm$ function. When $\beta_\pm > 0$, it is possible to have timelike circular orbits (TCOs) since both the energy and angular momentum of the timelike particle are well defined, 
but one can not have spacelike circular orbits, since the energy and angular momentum of the spacelike particle are not well defined. 
Likewise, when $\beta_\pm < 0$ the opposite occurs: it is not possible to have TCOs, but it is possible to have spacelike circular orbits. 

It is also possible to conclude that the transition of $\beta_\pm$ from positive to negative values, and vice-versa, is entirely continuous, providing that we can have circular orbits, \textit{i.e.} $C(r) \geqslant 0$.

\bibliography{Bibliography}

\begin{thebibliography}{10}

\bibitem{Schmidt:1963wkp}
M.~Schmidt, ``{3C 273 : A Star-Like Object with Large Red-Shift},'' {\em
  Nature}, vol.~197, no.~4872, p.~1040, 1963.

\bibitem{Begelman:1984mw}
M.~C. Begelman, R.~D. Blandford, and M.~J. Rees, ``{Theory of extragalactic
  radio sources},'' {\em Rev. Mod. Phys.}, vol.~56, pp.~255--351, 1984.

\bibitem{Kerr:1963ud}
R.~P. Kerr, ``{Gravitational field of a spinning mass as an example of
  algebraically special metrics},'' {\em Phys. Rev. Lett.}, vol.~11,
  pp.~237--238, 1963.

\bibitem{Syunyaev:1986zz}
R.~A. Syunyaev and N.~I. Shakura, ``{Disk Accretion onto a Weak Field Neutron
  Star - Boundary Layer Disk Luminosity Ratio},'' {\em Sov. Astron. Lett.},
  vol.~12, pp.~117--120, 1986.

\bibitem{hobson2006general}
M.~P. Hobson, G.~P. Efstathiou, and A.~N. Lasenby, {\em General relativity: an
  introduction for physicists}.
\newblock Cambridge University Press, 2006.

\bibitem{thorne2000gravitation}
K.~S. Thorne, C.~W. Misner, and J.~A. Wheeler, {\em Gravitation}.
\newblock Freeman, 2000.

\bibitem{Abbott:2016blz}
B.~P. Abbott {\em et~al.}, ``{Observation of Gravitational Waves from a Binary
  Black Hole Merger},'' {\em Phys. Rev. Lett.}, vol.~116, no.~6, p.~061102,
  2016.

\bibitem{LIGOScientific:2018mvr}
B.~P. Abbott {\em et~al.}, ``{GWTC-1: A Gravitational-Wave Transient Catalog of
  Compact Binary Mergers Observed by LIGO and Virgo during the First and Second
  Observing Runs},'' {\em Phys. Rev. X}, vol.~9, no.~3, p.~031040, 2019.

\bibitem{Akiyama:2019fyp}
K.~Akiyama {\em et~al.}, ``{First M87 Event Horizon Telescope Results. V.
  Physical Origin of the Asymmetric Ring},'' {\em Astrophys. J. Lett.},
  vol.~875, no.~1, p.~L5, 2019.

\bibitem{Akiyama:2019cqa}
K.~Akiyama {\em et~al.}, ``{First M87 Event Horizon Telescope Results. I. The
  Shadow of the Supermassive Black Hole},'' {\em Astrophys. J. Lett.},
  vol.~875, p.~L1, 2019.

\bibitem{Akiyama:2019eap}
K.~Akiyama {\em et~al.}, ``{First M87 Event Horizon Telescope Results. VI. The
  Shadow and Mass of the Central Black Hole},'' {\em Astrophys. J. Lett.},
  vol.~875, no.~1, p.~L6, 2019.

\bibitem{Akiyama:2021qum}
K.~Akiyama {\em et~al.}, ``{First M87 Event Horizon Telescope Results. VII.
  Polarization of the Ring},'' {\em Astrophys. J. Lett.}, vol.~910, no.~1,
  p.~L12, 2021.

\bibitem{Cunha:2017qtt}
P.~V.~P. Cunha, E.~Berti, and C.~A.~R. Herdeiro, ``{Light-Ring Stability for
  Ultracompact Objects},'' {\em Phys. Rev. Lett.}, vol.~119, no.~25, p.~251102,
  2017.

\bibitem{Cunha:2020azh}
P.~V.~P. Cunha and C.~A.~R. Herdeiro, ``{Stationary black holes and light
  rings},'' {\em Phys. Rev. Lett.}, vol.~124, no.~18, p.~181101, 2020.

\bibitem{Vieira:2017zau}
R.~S.~S. Vieira, W.~Klu\'zniak, and M.~Abramowicz, ``{Curvature dependence of
  relativistic epicyclic frequencies in static, axially symmetric
  spacetimes},'' {\em Phys. Rev. D}, vol.~95, no.~4, p.~044008, 2017.

\bibitem{SCHUNCK1998389}
F.~E. Schunck and E.~W. Mielke, ``Rotating boson star as an effective mass
  torus in general relativity,'' {\em Physics Letters A}, vol.~249, no.~5,
  pp.~389--394, 1998.

\bibitem{PhysRevD.56.762}
S.~Yoshida and Y.~Eriguchi, ``Rotating boson stars in general relativity,''
  {\em Phys. Rev. D}, vol.~56, pp.~762--771, Jul 1997.

\bibitem{Delgado:2016jxq}
J.~F.~M. Delgado, C.~A.~R. Herdeiro, E.~Radu, and H.~Runarsson, ``{Kerr-Newman
  black holes with scalar hair},'' {\em Phys. Lett. B}, vol.~761, pp.~234--241,
  2016.

\bibitem{Guerra:2019srj}
D.~Guerra, C.~F.~B. Macedo, and P.~Pani, ``{Axion boson stars},'' {\em JCAP},
  vol.~09, no.~09, p.~061, 2019.
\newblock [Erratum: JCAP 06, E01 (2020)].

\bibitem{Delgado:2020udb}
J.~F.~M. Delgado, C.~A.~R. Herdeiro, and E.~Radu, ``{Rotating Axion Boson
  Stars},'' {\em JCAP}, vol.~06, p.~037, 2020.

\bibitem{Brito:2015pxa}
R.~Brito, V.~Cardoso, C.~A.~R. Herdeiro, and E.~Radu, ``{Proca stars:
  Gravitating Bose\textendash{}Einstein condensates of massive spin 1
  particles},'' {\em Phys. Lett. B}, vol.~752, pp.~291--295, 2016.

\bibitem{Herdeiro:2017fhv}
C.~A.~R. Herdeiro, A.~M. Pombo, and E.~Radu, ``{Asymptotically flat scalar,
  Dirac and Proca stars: discrete vs. continuous families of solutions},'' {\em
  Phys. Lett. B}, vol.~773, pp.~654--662, 2017.

\bibitem{Minamitsuji:2018kof}
M.~Minamitsuji, ``{Vector boson star solutions with a quartic order
  self-interaction},'' {\em Phys. Rev. D}, vol.~97, no.~10, p.~104023, 2018.

\bibitem{Delgado:2020hwr}
J.~F.~M. Delgado, C.~A.~R. Herdeiro, and E.~Radu, ``{Kerr black holes with
  synchronised axionic hair},'' 12 2020.

\bibitem{Sotiriou:2014pfa}
T.~P. Sotiriou and S.-Y. Zhou, ``{Black hole hair in generalized scalar-tensor
  gravity: An explicit example},'' {\em Phys. Rev. D}, vol.~90, p.~124063,
  2014.

\bibitem{Delgado:2020rev}
J.~F.~M. Delgado, C.~A.~R. Herdeiro, and E.~Radu, ``{Spinning black holes in
  shift-symmetric Horndeski theory},'' {\em JHEP}, vol.~04, p.~180, 2020.

\bibitem{Carter:1970ea}
B.~Carter, ``{The commutation property of a stationary, axisymmetric system},''
  {\em Commun. Math. Phys.}, vol.~17, pp.~233--238, 1970.

\bibitem{Wald:1984rg}
R.~M. Wald, {\em {General Relativity}}.
\newblock Chicago, USA: Chicago Univ. Pr., 1984.

\bibitem{Collodel:2021gxu}
L.~G. Collodel, D.~D. Doneva, and S.~S. Yazadjiev, ``{Circular Orbit Structure
  and Thin Accretion Disks around Kerr Black Holes with Scalar Hair},'' {\em
  Astrophys. J.}, vol.~910, no.~1, p.~52, 2021.

\bibitem{Berti:2015itd}
E.~Berti {\em et~al.}, ``{Testing General Relativity with Present and Future
  Astrophysical Observations},'' {\em Class. Quant. Grav.}, vol.~32, p.~243001,
  2015.

\bibitem{Cao:2016zbh}
Z.~Cao, A.~Cardenas-Avendano, M.~Zhou, C.~Bambi, C.~A.~R. Herdeiro, and
  E.~Radu, ``{Iron K$\alpha$ line of boson stars},'' {\em JCAP}, vol.~10,
  p.~003, 2016.

\bibitem{Vincent:2015xta}
F.~H. Vincent, Z.~Meliani, P.~Grandclement, E.~Gourgoulhon, and O.~Straub,
  ``{Imaging a boson star at the Galactic center},'' {\em Class. Quant. Grav.},
  vol.~33, no.~10, p.~105015, 2016.

\bibitem{Cunha:2016bjh}
P.~V.~P. Cunha, J.~Grover, C.~Herdeiro, E.~Radu, H.~Runarsson, and A.~Wittig,
  ``{Chaotic lensing around boson stars and Kerr black holes with scalar
  hair},'' {\em Phys. Rev. D}, vol.~94, no.~10, p.~104023, 2016.

\bibitem{Ni:2016rhz}
Y.~Ni, M.~Zhou, A.~Cardenas-Avendano, C.~Bambi, C.~A.~R. Herdeiro, and E.~Radu,
  ``{Iron K$\alpha$ line of Kerr black holes with scalar hair},'' {\em JCAP},
  vol.~07, p.~049, 2016.

\bibitem{Cunha:2015yba}
P.~V.~P. Cunha, C.~A.~R. Herdeiro, E.~Radu, and H.~F. Runarsson, ``{Shadows of
  Kerr black holes with scalar hair},'' {\em Phys. Rev. Lett.}, vol.~115,
  no.~21, p.~211102, 2015.

\bibitem{Herdeiro:2021lwl}
C.~A.~R. Herdeiro, A.~M. Pombo, E.~Radu, P.~V.~P. Cunha, and N.~Sanchis-Gual,
  ``{The imitation game: Proca stars that can mimic the Schwarzschild
  shadow},'' {\em JCAP}, vol.~04, p.~051, 2021.

\bibitem{Herdeiro:2020kba}
C.~A.~R. Herdeiro, G.~Panotopoulos, and E.~Radu, ``{Tidal Love numbers of Proca
  stars},'' {\em JCAP}, vol.~08, p.~029, 2020.

\bibitem{Shen:2016acv}
T.~Shen, M.~Zhou, C.~Bambi, C.~A.~R. Herdeiro, and E.~Radu, ``{Iron K$\alpha$
  line of Proca stars},'' {\em JCAP}, vol.~08, p.~014, 2017.

\bibitem{Bryant:2021xdh}
A.~Bryant, H.~O. Silva, K.~Yagi, and K.~Glampedakis, ``{Eikonal quasinormal
  modes of black holes beyond general relativity. III. Scalar Gauss-Bonnet
  gravity},'' {\em Phys. Rev. D}, vol.~104, no.~4, p.~044051, 2021.

\bibitem{solver1}
W.~Sch\"{o}nauer and E.~Schnepf, ``Software considerations for the "black box";
  solver fidisol for partial differential equations,'' {\em ACM Trans. Math.
  Softw.}, vol.~13, pp.~333--349, dec 1987.

\bibitem{solver3}
W.~Schönauer and T.~Adolph, ``How we solve pdes,'' {\em Journal of
  Computational and Applied Mathematics}, vol.~131, no.~1–2, pp.~473 -- 492,
  2001.

\bibitem{Herdeiro:2015gia}
C.~Herdeiro and E.~Radu, ``{Construction and physical properties of Kerr black
  holes with scalar hair},'' {\em Class. Quant. Grav.}, vol.~32, no.~14,
  p.~144001, 2015.

\bibitem{PhysRev.172.1331}
D.~J. Kaup, ``Klein-gordon geon,'' {\em Phys. Rev.}, vol.~172, pp.~1331--1342,
  Aug 1968.

\bibitem{PhysRev.187.1767}
R.~Ruffini and S.~Bonazzola, ``Systems of self-gravitating particles in general
  relativity and the concept of an equation of state,'' {\em Phys. Rev.},
  vol.~187, pp.~1767--1783, Nov 1969.

\bibitem{Schunck:2003kk}
F.~E. Schunck and E.~W. Mielke, ``{General relativistic boson stars},'' {\em
  Class. Quant. Grav.}, vol.~20, pp.~R301--R356, 2003.

\bibitem{Herdeiro:2019mbz}
C.~Herdeiro, I.~Perapechka, E.~Radu, and Y.~Shnir, ``{Asymptotically flat
  spinning scalar, Dirac and Proca stars},'' {\em Phys. Lett. B}, vol.~797,
  p.~134845, 2019.

\bibitem{diCortona:2015ldu}
G.~Grilli~di Cortona, E.~Hardy, J.~Pardo~Vega, and G.~Villadoro, ``{The QCD
  axion, precisely},'' {\em JHEP}, vol.~01, p.~034, 2016.

\bibitem{jetzer1989charged}
P.~Jetzer and J.~J. Van~der Bij, ``Charged boson stars,'' {\em Physics Letters
  B}, vol.~227, no.~3-4, pp.~341--346, 1989.

\bibitem{Pugliese:2013gsa}
D.~Pugliese, H.~Quevedo, J.~A. Rueda~H., and R.~Ruffini, ``{On charged boson
  stars},'' {\em Phys. Rev. D}, vol.~88, p.~024053, 2013.

\bibitem{Brihaye:2009dxv2}
Y.~Brihaye, T.~Caebergs, and T.~Delsate, ``Charged-spinning-gravitating
  q-balls,'' {\em arXiv preprint arXiv:0907.0913}, 2009.

\bibitem{Collodel:2019ohy}
L.~G. Collodel, B.~Kleihaus, and J.~Kunz, ``{Structure of rotating charged
  boson stars},'' {\em Phys. Rev. D}, vol.~99, no.~10, p.~104076, 2019.

\bibitem{Santos:2020pmh}
N.~M. Santos, C.~L. Benone, L.~C.~B. Crispino, C.~A.~R. Herdeiro, and E.~Radu,
  ``{Black holes with synchronised Proca hair: linear clouds and fundamental
  non-linear solutions},'' {\em JHEP}, vol.~07, p.~010, 2020.

\bibitem{Herdeiro:2014goa}
C.~A.~R. Herdeiro and E.~Radu, ``{Kerr black holes with scalar hair},'' {\em
  Phys. Rev. Lett.}, vol.~112, p.~221101, 2014.

\bibitem{horndeski1974second}
G.~W. Horndeski, ``Second-order scalar-tensor field equations in a
  four-dimensional space,'' {\em International Journal of Theoretical Physics},
  vol.~10, no.~6, pp.~363--384, 1974.

\bibitem{Sotiriou:2013qea}
T.~P. Sotiriou and S.-Y. Zhou, ``{Black hole hair in generalized scalar-tensor
  gravity},'' {\em Phys. Rev. Lett.}, vol.~112, p.~251102, 2014.

\bibitem{Pugliese:2011py}
D.~Pugliese, H.~Quevedo, and R.~Ruffini, ``{Motion of charged test particles in
  Reissner-Nordstr\"om spacetime},'' {\em Phys. Rev. D}, vol.~83, p.~104052,
  2011.

\bibitem{Stuchlik:2002tj}
Z.~Stuchlik and S.~Hledik, ``{Properties of the Reissner-Nordstr\"om Spacetimes
  with a Nonzero Cosmological Constant},'' {\em Acta Phys. Slov.}, vol.~52,
  no.~5, pp.~363--407, 2002.

\bibitem{Stuchlik:2014yaa}
Z.~Stuchlik and J.~Schee, ``{Optical effects related to Keplerian discs
  orbiting Kehagias-Sfetsos naked singularities},'' {\em Class. Quant. Grav.},
  vol.~31, p.~195013, 2014.

\bibitem{Vieira:2013jga}
R.~S.~S. Vieira, J.~Schee, W.~Klu\'zniak, Z.~Stuchl\'\i{}k, and M.~Abramowicz,
  ``{Circular geodesics of naked singularities in the Kehagias-Sfetsos metric
  of Ho\v{r}ava\textquoteright{}s gravity},'' {\em Phys. Rev. D}, vol.~90,
  no.~2, p.~024035, 2014.

\bibitem{Collodel:2021jwi}
L.~G. Collodel, D.~D. Doneva, and S.~S. Yazadjiev, ``{Equatorial
  extreme-mass-ratio inspirals in Kerr black holes with scalar hair
  spacetimes},'' {\em Phys. Rev. D}, vol.~105, no.~4, p.~044036, 2022.

\end{thebibliography}
\bibliographystyle{ieeetr}

\end{document}